\documentclass[fleqn,usenatbib]{mnras}

\usepackage{newtxtext,newtxmath}

\usepackage[T1]{fontenc}

\DeclareRobustCommand{\VAN}[3]{#2}
\let\VANthebibliography\thebibliography
\def\thebibliography{\DeclareRobustCommand{\VAN}[3]{##3}\VANthebibliography}

\usepackage{graphicx}	
\usepackage{amsmath}	
\usepackage{hyperref}
\hypersetup{colorlinks=True,citecolor=blue}
\usepackage{xcolor}
\usepackage{fontawesome5}
\usepackage{pdflscape}
\usepackage{makecell}

\usepackage{color}

\def\ms{\hbox{\,m\,s$^{-1}$}}         
\def\m2s2{\hbox{\,m$^{2}$\,s$^{-2}$}} 
\def\sini{\hbox{sin\,$i$}}      
\def\Mearth{\hbox{$\mathrm{M}_{\oplus}$}}             
\def\Msun{\hbox{$\mathrm{M}_{\odot}$}}             

\def\S/N{$S/N_{cont}$}

\def\prec{\emph{precision}}
\def\acc{\emph{accuracy}}
\def\score{\emph{maximum error}}

\def\Halpha{H\hspace{0.1ex}$\alpha$ }

\def\K3{K$_3$ } 
\def\K2{K$_2$ }
\def\K2v{K$_{2V}$ }
\def\K2r{K$_{2R}$ }

\title{Measuring masses of transiting Earth-like planets: first insights from the PLATO RV Data Challenge}

\author[M. Cretignier]{
M. Cretignier$^{1}$\thanks{E-mail: michael.cretignier@physics.ox.ac.uk},
N. C. Hara$^{2}$, 
S. Aigrain$^{1}$, 
S. Sulis$^{2}$, 
F. Bouchy$^{3}$, 
S. T. Hodgkin$^{4}$,
J. Blanco-Pozo$^{5,6,7}$, \newauthor
A. Collier-Cameron$^{8}$,  
M. Damasso$^{9}$, 
E. C. Gonçalves$^{10}$, 
L. Malavolta$^{11,12}$, 
G. Mantovan$^{12}$, \newauthor
D. Nardiello$^{11,12}$, 
M. Perger$^{5,6}$, 
I. Ribas$^{5,6}$, 
Y. Zhao$^{13}$,
X. Dumusque$^{3}$, 
A. Mortier$^{14}$, 
K. Al Moulla$^{10}$, \newauthor
S. C. C. Barros$^{10}$, 
A. Santerne$^{2,18}$,
N. C. Santos$^{10,15}$, 
C. San Nicolas Martinez$^{10}$, 
C. Watson$^{16}$,
T. Wilson$^{17}$, \newauthor
S. Udry$^{3}$ 
and the PLATO Stellar Variability Working Group
\\
$^{1}$Department of Physics, University of Oxford, OX13RH Oxford, UK \\
$^{2}$Aix Marseille Univ, CNRS, CNES, LAM, Marseille, France\\
$^{3}$Astronomy Department of the University of Geneva, 51 ch. de Pegasi, 1290 Versoix, Switzerland\\
$^{4}$Institute of Astronomy, Madingley Road, Cambridge, CB3 0HA, UK  \\
$^{5}$Institut de Ciències de l'Espai (ICE, CSIC), Campus UAB, Carrer de Can Magrans s/n, 08193 Bellaterra, Barcelona, Spain\\
$^{6}$Institut d’Estudis
Espacials de Catalunya (IEEC), Edifici RDIT, Campus UPC, 08860 Castelldefels,
Barcelona, Spain \\
$^{7}$Facultat de Física, Universitat de Barcelona (UB), Martí Franquès 1, 08028, Barcelona, Spain\\
$^{8}$SUPA School of Physics and Astronomy, University of St Andrews, North Haugh, St Andrews KY16 9SS, UK\\
$^{9}$INAF – Osservatorio Astrofisico di Torino, Via Osservatorio 20, 10025 Pino Torinese, Italy\\
$^{10}$Instituto de Astrofísica e Ciências do Espaço, Universidade do Porto, CAUP, Rua das Estrelas, 4150-762 Porto, Portugal \\
$^{11}$Dipartimento di Fisica e Astronomia 'Galileo Galilei' -- Universit\'a degli Studi di Padova, Vicolo dell’Osservatorio 3, IT-35122, Padova, Italy \\
$^{12}$Centro di Ateneo di Studi e Attivit\`a Spaziali ``G. Colombo'' -- Universit\`a degli Studi di Padova, Via Venezia 15, IT-35131, Padova, Italy \\
$^{13}$Department of Astronomy, University of Texas at Austin, 2515 Speedway, Austin, TX 78712, USA\\
$^{14}$School of Physics \& Astronomy, University of Birmingham, Edgbaston, Birmingham, B15 2TT, UK\\
$^{15}$Departamento de F\'isica e Astronomia, Faculdade de Ci\^encias, Universidade do Porto, Rua do Campo Alegre, 4169-007 Porto, Portugal \\
$^{16}$Astrophysics Research Centre, School of Mathematics and Physics, Queen’s University Belfast, Belfast, BT7 1NN, UK \\
$^{17}$Department of Physics, University of Warwick, Coventry, CV4 7AL, UK \\
$^{18}$Univ. Grenoble Alpes, CNRS, IPAG, 38000 Grenoble, France \\
}

\date{Accepted XXXX. Received XXXX; in original form XXXX}

\pubyear{2026}

\begin{document}
\label{firstpage}
\pagerange{\pageref{firstpage}--\pageref{lastpage}}
\maketitle

\begin{abstract}
The PLATO Stellar Variability Mitigation Working Group designed a radial-velocity (RV) data challenge to assess the precision and accuracy achievable on RV semi-amplitudes in the context of the future PLATO follow-up. 
Using Sun-as-a-star observations obtained with HARPS-N, we generated three multi-planet systems with realistic sampling, an effective signal-to-noise ratio of $S/N \sim 450$, an exposure time of $T_{\mathrm{exp}} = 15$ minutes, and a total of $N \sim 1000$ observations. The resulting data set is publicly available and can serve as a benchmark for the community to test and compare RV analysis methods. Eight independent teams participated in the challenge, aiming to recover the semi-amplitudes of both transiting and non-transiting planets. A meta-analysis of the submitted results shows that, with current data analysis methods, HARPS-N-like stability and the observing strategy studied in this paper, achieving a precision on semi-amplitudes better than 10\% for habitable-zone signals around a solar-type star is challenging, and is typically limited to planets with masses of $\sim 8$–$10$\,$\Mearth$. This performance is expected to improve for K dwarfs, where planets down to $\sim 2$–$5$ \,$\Mearth$ may be characterized at a similar level of precision. Finally, based on the results of this challenge and the occurrence of false positives, we derive an empirical detection threshold for RV surveys, which can be used to distinguish robust planetary detections from candidates.
\end{abstract}




\section{Introduction}
\label{sec:introduction}

The transit method is one of the most efficient and prolific techniques for detecting exoplanets today. Previous space missions such as Kepler \citep[launched in 2009,][]{Koch(2010),Borucki(2010)} and TESS \citep[launched in 2018,][]{Ricker(2015)} have considerably improved our understanding of planetary demographics, particularly for short-period planets \citep{Petigura(2013),Batalha(2014),Burke(2015),Bryson(2020), Royer(2025), Cui(2026)}.

The transiting configuration naturally biases detections towards short-period planets. However, thanks to their large fields of view monitoring thousands of stars simultaneously, both missions have produced numerous and prominent detections. This success is largely due to the fact that the signature of a transiting planet is highly orthogonal 
to most stellar photometric variability signals, such as granulation and rotational modulation. However, these two missions have shortcomings. Most stars in the Kepler field are faint, which makes the mass measurements difficult. Whereas TESS targets are brighter, the observation strategy focuses on short period planets. 



PLATO (PLAnetary Transits and Oscillations of stars) is an upcoming space-based photometric mission designed to observe bright stars, such that follow-up observations are possible, and its observation strategy is designed to enable the detection of Earth analogs  
\citep{Rauer(2014),Rauer(2024),Rauer(2025)}. Indeed, the survey strategy is broadly similar to that of Kepler, with a long-duration pointing towards a southern field observed for about at least two years. This baseline enables the detection of at least two transits of Earth-like planets, or potentially single-transit (mono-transit) events. The nominal mission is expected to continue for an additional two years, either by switching to a northern field or by extending the southern observations, resulting in a total time baseline comparable to Kepler’s four-year monitoring. Furthermore, its photometric precision is expected to be close of Kepler, with an anticipated noise level of about 20–30 ppm on hourly timescales for bright targets \citep{Janssen(2024)}. Unlike Kepler, however, PLATO will primarily target brighter stars \citep{Cabrera(2026),Montalto(2026)}, a key advantage for enabling precise ground-based radial-velocity follow-up and mass characterization of the detected planets.

 The radial velocity (RV) follow-up is an integral part of the mission and aims at measuring the mass of the exoplanets detected by PLATO, including Earth analogs. Combined with the radius estimate, this will enable a mean density measurement, and yield constraints on the interior structure. Breaking the degeneracy between several interior compositions requires a high accuracy on the density, ideally $\sim$10\% for terrestrial planets \citep{plotnykov2024}, which translates to stringent requirements on the measurement of RV semi-amplitude, as seen below. 



\subsection{The PLATO mission requirements on mass measurements}
\label{sec:requirements}

The PLATO mission requirements R-SCI-Lo-15\footnote{\url{https://platomission.com/2018/05/19/top-scientific-requirements}} reads: \emph{PLATO shall observe stars that are bright enough to allow for a determination through radial velocity measurements of the mass of a terrestrial planet orbiting a G0V star with an accuracy of 10\% or better.
“Terrestrial planet” is defined as having R<2 $R_{\oplus}$ and M<10 $M_{\oplus}$} -- a goal that remains challenging for current RV capabilities.

In principle, the recent ultra-stable spectrographs are capable of reaching such a precision with sub \ms{} stability.
These include ESPRESSO \citep{Pepe(2014b)}, NEID \citep{Schwab(2018)}, EXPRES \citep{Jurgenson(2016)}, HARPS-N \citep{Cosentino(2012)}, MAROON-X \citep{Seifahrt(2018)} and KPF \citep{Gibson(2016)}, building upon the legacy of instruments such as HARPS \citep{Mayor(2003)}. For instance, ESPRESSO has demonstrated an intra-night precision of the order of 10 cm/s \citep{Pepe(2021),Figueira(2025)}. Unfortunately, at these precision levels, stellar activity dominates the RV signal, significantly complicating the detection of Earth-like planets around Sun-like stars \citep[see e.g.][]{Meunier(2020), Crass(2021), Hara(2023b)}. 

The challenge in RVs is greater than in photometry because the planetary signal is not a so-called gated signal (i.e., a component that appears and disappears over time), but instead is continuously superposed with all other contributions. This has important implications: for instance, machine-learning algorithms cannot be trained on subsets of the data that are “free” of the planetary signal, but also it means that any incomplete or inaccurate model will always interact with the planetary signals in unpredictable ways. In addition, the contaminating signals are more numerous, often larger in amplitude, and more problematic in the time and frequency domains \citep{Meunier(2015),Fischer(2016),Cretignier(2021),Crass(2021),Hara(2023b)}. 

Some attempts have been made to evaluate the RV limitations based on simulations \citep{Meunier(2019b),Meunier(2020),Meunier(2020c)}, but the results are subject to caution. On the one hand, the simulations might not be realistic enough, on the other hand the data analysis methods used in these studies might be too simplistic. 
Far from being discouraging, they motivate further intensive investigations based on real data sets and gathering more independent experts: ultimately the organisation of an RV challenge.   

In this paper, we aim to assess whether the PLATO mass-precision requirements can be fulfilled with current methods and the extent to which they can be met across the ($K,P$) parameter space. We have therefore designed an RV challenge involving several competitors and collaborators within the PLATO community, that used some state-of-the-art methodologies currently available. This challenge aims to mimic as closely as possible a PLATO follow-up. One of its most important aspects is that it makes use of real data, where planets are injected at the spectral level. 

In Sect.~\ref{sec:plato_challenge}, we introduce the main concept of the challenge. We begin by reviewing previous challenges (Sect.~\ref{sec:archive_challenge}), followed by a description of the design of the present challenge, including the injected planetary systems and the datasets used (Sect.~\ref{sec:design}). The participating teams and their methods are then presented in Sect.~\ref{sec:participants}. The results are described in Sect.~\ref{sec:results}, including the definition of the scoring metrics (Sect.~\ref{sec:secmetric}), the performance on the different datasets (Sect.~\ref{sec:result_individual}), and the overall ranking of the methods (Sect.~\ref{sec:ranking}). We then discuss the broader implications for the RV field (Sect.~\ref{sec:rvfield}), before concluding in Sect.~\ref{sec:conclusion}.

\section{A PLATO RV challenge}
\label{sec:plato_challenge}

\subsection{Previous RV challenges}
\label{sec:archive_challenge}

Few RV challenges have been conducted over the past decade. The main advantage of such community exercises is that they allow different methods to be tested on identical datasets. This enables a meaningful benchmarking of the field, provides a snapshot of the current state of the art, and allows the performance of the various approaches to be assessed within a consistent framework.

The first RV challenge was organized by \citet{Dumusque(2017)} using simulations as well as real HARPS data sets 
in which planetary signals were injected directly into the $RV(t)$ time series. Although realistic, the simulations were not capturing all the intricacies of stellar variability. Indeed, a linear fit of activity indicators worked extremely well to remove stellar activity signals in simulated systems, but performed worse in practice. Furthermore, this challenge was unfortunately limited to high-level data products from the data reduction software (DRS), namely the extracted RV time series and common activity proxies. In recent years, however, several studies have suggested that significant improvements can be achieved by working directly in the velocity domain \citep{Collier(2021),Beurs(2020),Zhao(2020),Zhao(2022b),Cretignier(2022),Yu(2024),Klein(2025)}, in the wavelength domain \citep{Cretignier(2021),Sharaf(2023),zhao(2024)} or by combining multiple RV time series derived from different spectral line selections \citep{Lienhard(2022),Dravins(2024),Almoulla(2024)} or obtained with different extraction methods \citep{Zechmeister(2018),Silva(2022)}.

More recently, \citet{Zhao_challenge} organized another challenge based on four EXPRES data sets, comparing different methodologies. In this case, data products at multiple processing levels were made available to participants, offering a broader and more up-to-date perspective on the field. However, the conclusions were less definitive. First, the metric used to compare methods is the root mean square of the corrected RVs, which might favor methods that are overfitting \citep{Hara(2023b)}. Second, no participant consistently outperformed the others across all four data sets, which complicates any attempt at establishing a robust ranking metric. 

Although it is well known that the true planetary content of real data sets can never be fully known since low undetected signals may always exists \footnote{Exceptions due to stability consideration may exists in very few specific cases \citep[see e.g][]{Zhao(2021)}}, the absence of injected signals in any of the data sets represents a limitation of the analysis. Injecting realistic planetary signals at the low-level data stage is technically challenging — and propagating those signals consistently through to high-level products is even more complex. However, we are aware that such an attempt is in progress (Zhao et al. in prep.).

\subsection{Design of the PLATO challenge}
\label{sec:design}

We designed the PLATO RV challenge to closely mimic the conditions expected for the RV follow-up of PLATO targets, implementing some recommendations of \cite{Hara(2023b)} on the design of RV challenges. To achieve this, several high-level design choices were made.

First, the challenge exclusively relies  on real data sets. While simulated data allow complete control over the ground truth, they inevitably lack certain complexities present in real observations, such as instrumental systematics and subtle correlated noise sources. These effects remain part of the real problem and should not be underestimated.

Second, planetary signals corresponding to transiting planets are injected at the low-level data stage on the spectra, rather than directly into derived RV time series. This ensures that injected signals propagate consistently through the full reduction and analysis chain.

Third, to place the RV challenge within PLATO’s core objective of detecting Earth-mass planets in the habitable zone (HZ) of Sun-like stars, we ensured that at least one transiting exoplanet is always located in the Sun’s HZ. This guarantees that the target star would be considered a high-priority candidate for follow-up within the Ground-based Observation Program (GOP) of PLATO. 

Finally, additional planetary signals from non-transiting planets are injected. This enables the methods to be evaluated under realistic conditions, where both known (transiting) and unknown (non-transiting) signals coexist. As a reminder, in the Solar System no more than three planets can be seen transiting from any given vantage point in the Universe \citep{Wells(2018)}. This is important as the number of measurements needed to estimate the mass at a given precision of transiting exoplanets increases if non-transiting companions are present \citep{he2021}.  
 Also, the detection of non-transiting planets is a more stringent test of RV methods. While it seems possible for a data analysis method to retrieve by chance a small semi-amplitude, close to the one of an injected planet, the odds of retrieving by chance the correct period of the injected signal are very low. 

\subsubsection{Injected planetary system }
\label{sec:planetary}

The parameter space of multi-Keplerian systems is extremely large. 
In this challenge, we therefore restricted ourselves to the production of three multi-planetary data sets and limited the analysis to purely circular orbits. Without loss of generality, the term Keplerian will be used hereafter, even if it should be understood as a \emph{sinusoid}, rather than a strict Keplerian. All the participants were aware of the null eccentricity hypothesis and therefore removed it from their Keplerian fit adjustment. This 
is certainly one of the least realistic aspects of the challenge, since in this configuration only the semi-amplitudes of the transiting signals need to be adjusted. However, as shown in \cite{shenturner2008} and \cite{baluev2015}, it is not significantly more difficult to detect eccentric planets than circular ones up to $e \approx 0.7$. This is also the limit from which computational schemes might struggle to estimate eccentricities \citep{hara2019}. There is no clear evidence that such eccentricies are common for low mass planets.

Transit measurements are known to yield far more precise constraints on phase $T_{0}$ and orbital period $P$ than RV observations. In practice, this precision would differ from system to system, adding an additional layer of complexity to the challenge analysis — particularly given the limited number of data sets we generated. Therefore, for all the transiting planets, we provided the exact orbital periods and transit phases such that all the participants set their priors on them to an arbitrary large precision. 
The performance of the teams will be solely evaluated based on their ability to retrieve the semi-amplitudes of the transiting planets, and find the transiting signals. 

\begin{figure}
	
	\centering
	\includegraphics[width=8.5cm]{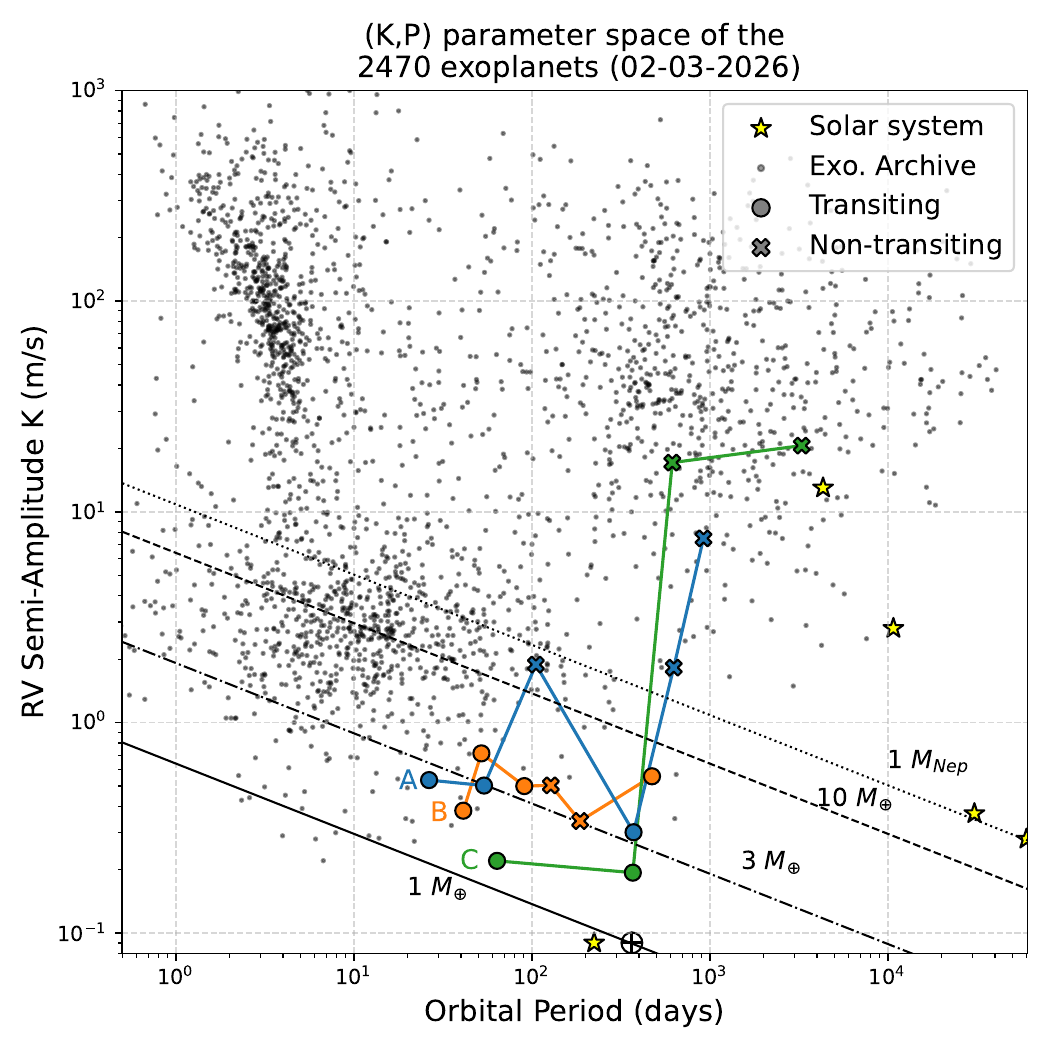}
	\caption{Illustration in the ($K$,$P$) parameter space of the three multi-planetary systems injected in the HARPS-N solar data: A (blue), B (orange) and C (green). Transiting planets are depicted with a round marker (see legend). All the transiting planets in this work were smaller than $7\Mearth$. Four isomasses lines are indicated. The 2470 exoplanets with a measured semi-amplitude according to NASA exoplanet archive are represented as a comparison.}
	\label{fig:KP_systems}
 
\end{figure}

Based on Doppler shifts injection tests performed in Appendix \ref{app:injection}, we restricted the values of the K semi-amplitudes below $\sim 25$ m/s for the injected planets. Above this threshold, the Doppler wavelength shift interpolation used to inject the planets in the spectra may lose their accurate validity to a degree that is critical for the RV challenge.

The three injected multi-planetary systems are represented in the $(K, P)$ parameter space in Fig.\ref{fig:KP_systems} and are hereafter referred to as systems A, B, and C. The Keplerian parameters are indicated in Table.\ref{tab:keplerians}. All terrestrial planets range between 2 to 7 \Mearth, that are already challenging planets to characterize today. 

For all the systems, we always assumed the solar properties for the host star, including the HZ limits even if we extrapolated the results to cooler type stars later (see Sect.\ref{sec:kdwarf}).

\paragraph{System A}


\begin{table}
 \caption{Summary table of the three multiplanetary systems A,B and C injected in the PLATO RV challenge. Information available to the participants is indicated in bold. The orbital period $P$, RV semi-amplitude $K$, planetary mass $m_p$ and transit phase $T_c$ are indicated. The type column indicate to which category a planet belongs (T=transit, M=mono-transit). Note that both the system A and C have their HZ signal close to 365 days which implies an incomplete phase coverage due to season gaps. The semi-amplitude $K_{\text{raw}}$ obtained by fitting the system on the data without the planet injected is also indicated with the relative amplitude with injected signals. }
 \label{tab:keplerians}
 \begin{tabular}{lllllcl}
  \hline
 Syst. & $P$ & $K$ & $m_p$ & $T_c$ & Type  & $K_{\text{raw}}$ \\
  & [days] & [m/s] & [$\Mearth$] & [BJD] &   & [m/s] (\%)\\
  \hline
A  &   \textbf{26.47} &  0.53 &  2.49 & \textbf{57152.2818} & T  & 0.11 (21\%) \\
A  &   \textbf{53.90} &  0.50 &  2.97 & \textbf{57163.1609} & T  & 0.15 (30\%) \\
A  &   105.5 &  1.88 &  13.9 & 57166.2604 & - & 0.11 (6\%)\\
A  &   \textbf{373.0} &  0.30 &  3.4 &  \textbf{57142.0194} & T & 0.12 (40\%)\\
A  &   627.3 &  1.82 &  24.4 &  57186.9142 & - & 0.27 (15\%)\\
A  &   921.5 &  7.48 &  114 & 57000.0000 & - & 0.92 (12\%) \\  \hline
B  &   \textbf{41.21} &  0.38 &  2.06 &  \textbf{54971.5052} & T & 0.05 (13\%)\\
B  &   \textbf{52.10} &  0.71 &  4.17 &  \textbf{54971.1748} & T & 0.11 (15\%)\\
B  &   \textbf{90.74} &  0.50 &  3.51 &  \textbf{54981.4550} & T & 0.20 (40\%)\\
B  &   128.0 &  0.50 &  3.97 &  54987.1590 & - & 0.20 (40\%) \\
B  &   186.8 &  0.34 &  3.05 &  54964.6487 & - & 0.15 (44\%) \\
B  &   473.5 &  0.56 &  6.78 &  \textbf{55120.2901} & M & 0.29 (52\%) \\  \hline
C  &   \textbf{63.79} &  0.22 &  1.38 & \textbf{59640.1579} & T & 0.13 (59\%) \\
C  &   \textbf{369.5} &  0.19 &  2.18 & \textbf{59646.0919} & T  & 0.35 (184\%)\\
C  &   615.9 &  17.1 &  228 & 59641.5763 & -  & 0.80 (5\%)\\
C  &   3286 &  20.6 &  480 & 59656.0921 & -  & 2.86 (14\%) \\  \hline

  \hline
 \end{tabular}
\end{table}

System A is a hierarchical six-planet system composed of three transiting terrestrial planets, two non-transiting Neptune-sized planets, and one non-transiting Saturn-mass planet.

The system architecture is inspired by the HIP 41378 system, which hosts both transiting and non-transiting planets spanning a wide range of orbital periods, and thus provides a realistic benchmark for complex planetary configurations \citep[e.g][]{Vanderburg(2016b), Santerne(2019),Grouffal(2025)}. To adapt this architecture to our study, we rescaled the orbital periods by a factor of 1.7 for all planets, except for the fourth planet.

As a result, the orbital period of the innermost planet is close to the solar rotation period around 27 days, providing a stringent test for disentangling planetary signals from stellar activity due to increased signal crosstalk in both photometric and RV time series.

The third planet is non-transiting, similarly to HIP 41378 g, and contributes only through dynamical and RV signatures. The outermost planet is also non-transiting and induces a long-period dominant signal in the RV data.

The fourth planet was adjusted to transit and to lie within the HZ of a solar-type (G2V) star, corresponding to orbital periods between 338 and 586 days. To satisfy this constraint while preserving the system hierarchy, a specific scaling factor of 1.36 was applied to its orbital period. This planet is designed to mimic a high-priority PLATO candidate suitable for follow-up observations.

Planetary radii were scaled by factors between 2 and 2.5 in order to control the resulting RV semi-amplitudes. A relation $m_p\propto R_p^{3.7}$ was assumed to convert radii to masses following \citep{Zeng(2016)}. By doing so, we ensured that all injected signals remain below 1 m/s, consistent with the low-amplitude regime expected for PLATO targets.

\paragraph{System B} 


System B is a compact six-planet system composed exclusively of rocky planets, inspired by the Kepler-11 system \citep{Lissauer(2011)}. Such systems are representative of tightly packed inner planetary systems commonly observed by transit surveys, making them a relevant benchmark for testing detection performance in crowded configurations.

The system includes four transiting planets (the three innermost and the outermost) and two non-transiting planets. The outermost planet is located within the HZ. In this configuration, the habitable-zone planet produces a mono-transit event. Only the transit epoch was provided, while the orbital period was left unconstrained\footnote{A weak prior was given, indicating that the period was expected to lie between 200 and 500 days.}, mimicking realistic observational limitations.

The planetary radii were scaled by factors of 1.6, 2.1, 2.4, 2.5, 1.9, and 2.1, while all orbital periods were increased by a factor of 4.  This makes all the RV semi-amplitudes below 70 cm/s, placing the system in a regime where stellar activity dominates the RV signal, thereby significantly complicating planet detection and characterization.

\paragraph{System C} 




System C is a four-planet system composed of two transiting terrestrial planets and two long-period non-transiting giant planets.

The system architecture was inspired by the HD 110067 system \citep{Luque(2023)}, with modified orbital periods and two planets removed to simplify the configuration (planets d, e, f, g only). Their orbital periods were rescaled by factors of 7, 12, 15, and 60. Planetary radii were kept unchanged, except for the innermost planet, whose radius was reduced by a factor of 2.

The two inner planets are terrestrial and transiting. The second planet is located within the HZ to mimic a high-priority PLATO target for follow-up observations. The two outer planets are long-period, non-transiting giant planets.

Despite the reduced number of injected planets compared to Systems A and B, the terrestrial signals in System C are among the most challenging to detect, with RV semi-amplitudes as low as 20 cm/s. 

The outermost giant planet has an orbital period close to the 11-year solar magnetic cycle (Schwabe cycle), introducing additional difficulty in disentangling long-period planetary signals from stellar magnetic variability.

\subsubsection{HARPS-N solar data set}
\label{sec:data set}

To create the RV challenge, we used the HARPS-N solar telescope data set \citep{Dumusque(2015), Dumusque(2021)}. HARPS-N has been observing the Sun daily since 2015 with 5-minute integration times continuously during the day. We used the data release corresponding to \citet{Dumusque(2021)}, although a more recent release became available this year \citep{Dumusque(2026)}. The baseline of the observations span eight years between July 2015 (BJD = 2,457,220) up to October 2023 (BJD = 2,460,238).

The use of HARPS-N solar data offers several advantages compared to archival stellar RV databases. A key benefit is that, for the Sun, the planetary signals of the Solar System can be removed, thereby providing a well-defined ground truth. In addition, the exceptional temporal sampling of the solar observations enables flexible control of the cadence, allowing it to be reshaped in a realistic manner to closely match the expected sampling of the PLATO RV follow-up.

However, several important caveats must be considered when extrapolating solar results to stellar observations. 

First, although difficult to quantify, it should be noted that some participants may have previously used or worked with the HARPS-N solar dataset. This potentially affects the strictly “blind” nature of the challenge\footnote{The use of the same dataset across multiple realizations also partially compromises the blindness of the challenge, as the absence or presence of signals between datasets may hint at injected planets. This is in particular the case for the giant ones.}, and may therefore introduce a slight imbalance between participants. This limitation is arguably an inherent trade-off when relying on a dataset of such importance, which has been widely used in the literature over the past years.

The datasets used in the challenge have an average visible bandpass\footnote{On HARPS-N, the average SNR of the visual bandpass equals to 83\% of the one at 5500\AA.} signal-to-noise ratio of $SNR = 450\pm40$. The impact of lower S/N on the results was not explored, although such an analysis would be valuable and could be performed through the injection of photon noise.

Third, because the Earth orbits the Sun, Sun-as-a-star data sets are less sensitive to some instrumental systematics that are fixed in the laboratory rest frame. When observing the Sun, the barycentric Earth radial velocity (BERV) span is limited to approximately $\sim$2 km/s (due to the Earth’s orbital eccentricity), whereas for typical stellar observations the BERV span can reach dozens of km/s. Given that the typical photospheric line width of a slow rotating star is $\sim$6 km/s for an HARPS-like instrumental resolution, the instrumental systematics in solar observations remain effectively “frozen” with respect to the line profile, whereas in stellar observations they are swept across the spectrum over time. This difference alters how instrumental effects couple to the spectral lines \citep[see e.g.][]{Silva(2025)}.

\begin{figure}
	
	\centering
	\includegraphics[width=8.5cm]{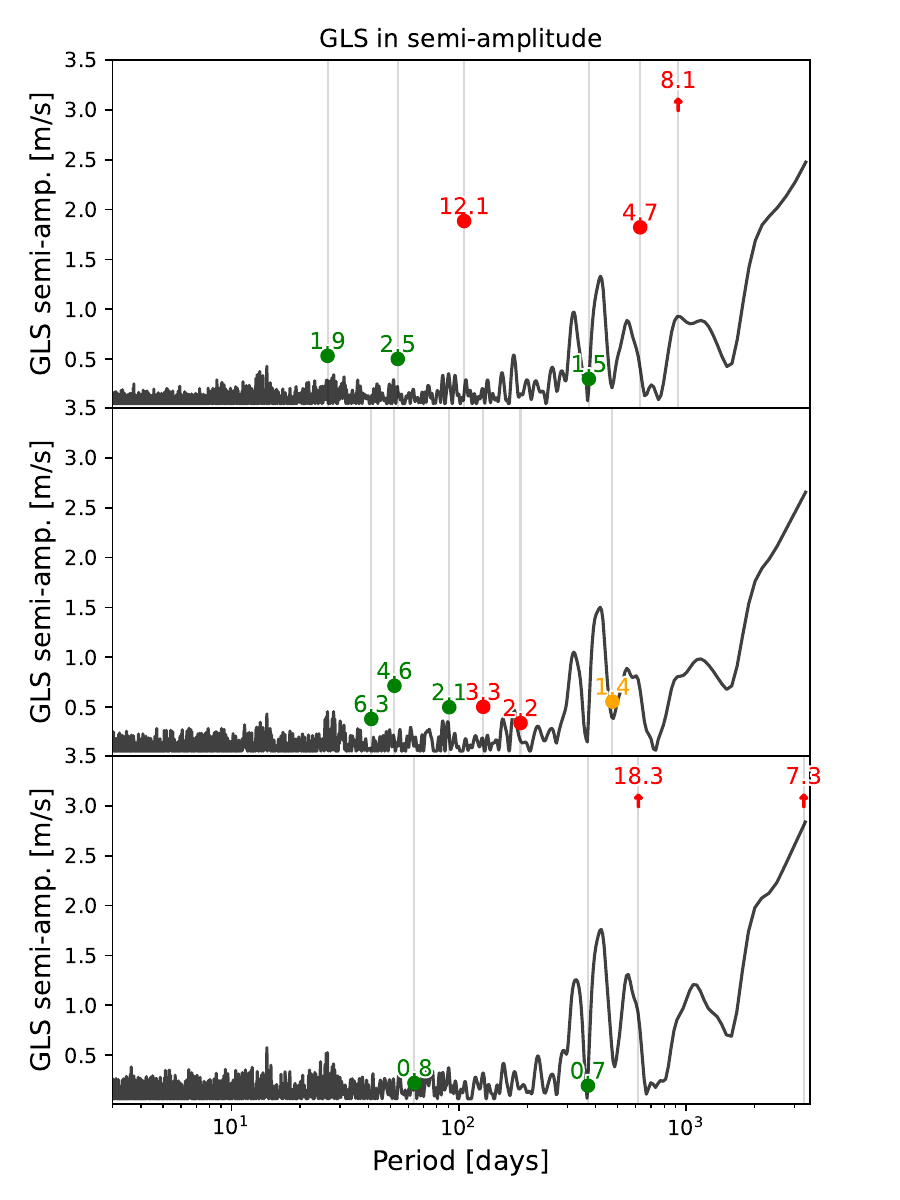}
	\caption{Illustration of the GLS periodogram for the three data set A1050 (top panel), B1300 (middle panel) and C800 (bottom panel) without the planets injected. Although the GLS is traditionally represented in terms of power, we display it here in $K$ semi-amplitude sinusoids scaling to facilitate a direct comparison with the semi-amplitudes of the injected signals. Transiting planets are shown in green (mono-transit in orange), non-transiting one in red. Signals larger than 3 m/s are just truncated and their marker is replaced by an upper arrow. We also  indicate above each marker the ratio of the planetary amplitude with the native GLS power at the same period.}
	\label{FigGLS}
\end{figure}

Last but not least, the solar telescope follows a different optical path than standard stellar observations. On the one hand, this removes some sources of systematics related to telescope guiding \citep[see e.g. the recently unperfect fibre scrambling identified on HARPS-N in][]{Anna(2025)}. On the other hand, the disk-resolved nature of the Sun introduces additional complications, particularly differential extinction across the solar disk. Although this effect has been calibrated at the RV level through an analytical correction term $RV_{\mathrm{corr}}(t)$ in \citet{Collier(2019)}, its correction at the spectral level has never been fully demonstrated or implemented. As a result, a simple Doppler shift by the measured $RV(t)$ is not strictly sufficient to correct the spectra.

Aware of this issue, we limited this effect doing the following selection of the spectra. First of all, we only selected spectra having a good quality flag. Bad data are usually identified as outliers in the daily magnitude vs. airmass relation which is expected to be linear \citep{Almoulla(2023)} and as such diagnose the periods in time when clouds occulted the Sun. We restricted the effect of the differential extinction by only selecting solar observation close to the meridian with an airmass smaller than $Z=1.3$. At such airmass, the differential extinction effect is lower than 10 cm/s according to the $RV_{\text{corr}}(t)$ vector. Note that this choice also recreates a realistic season length sampling since the Sun in winter is not going below an airmass $Z=1.3$.

\begin{figure*}
	\centering
	\includegraphics[width=18cm]{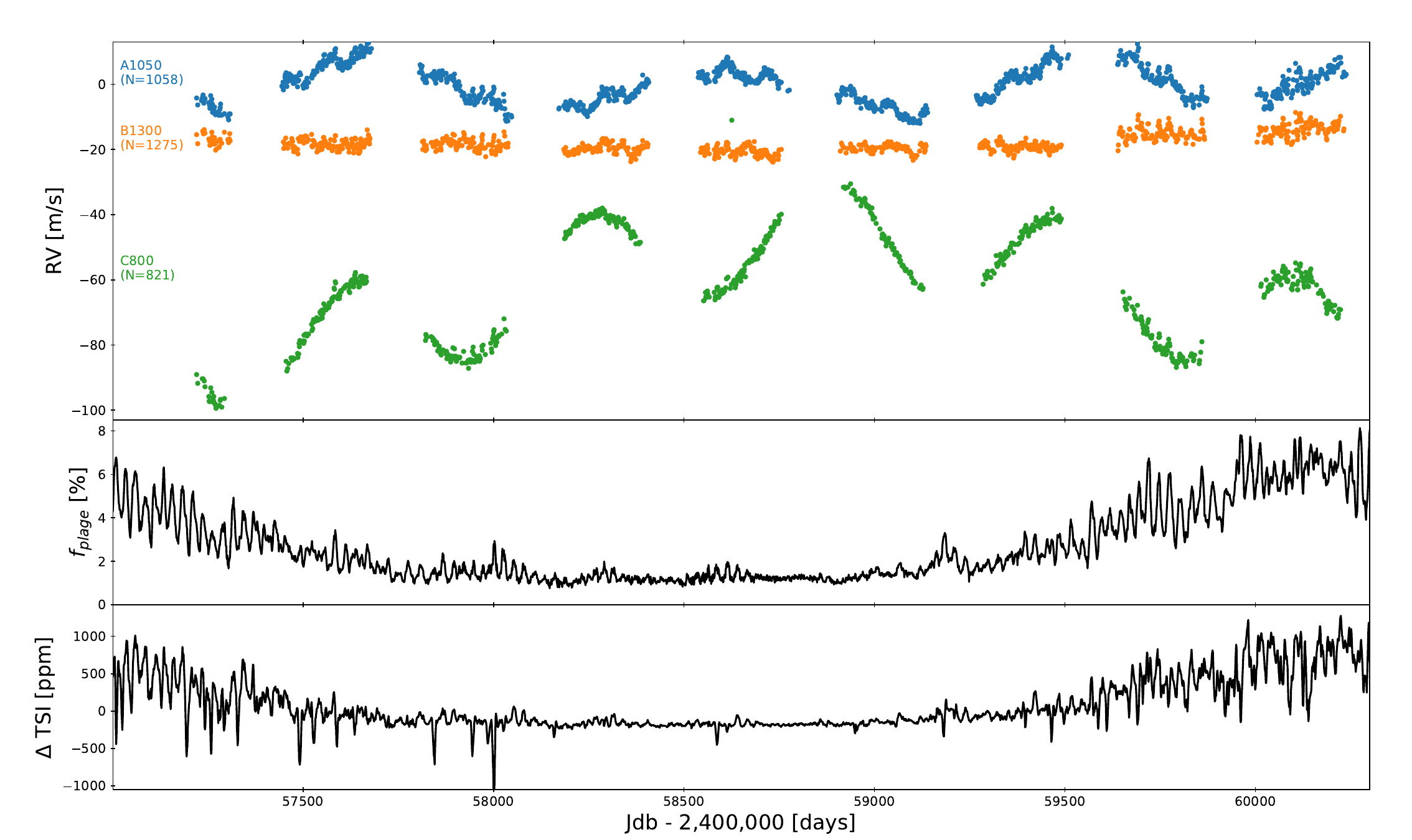}
	\caption{Representation of the main time-series provided for the RV challenge. \textbf{Top panel:} RV time-series (YV0) of the A1300 (blue), B1050 (orange) and C800 (green) data sets. Arbitrary vertical offset were applied for graphical considerations. \textbf{Middle panel:} The plage filling factor as extracted from the MgII index and converted here in plage filling factor using the reference time-series of \citet{Cretignier(2024)}. Bottom panel: Total solar irradiance converted here from usual $W/m^2$ units to \emph{ppm} by using solar minimum flux level as reference.}
	\label{fig:data sets}
\end{figure*}

Each day, we randomly selected three consecutive observations that satisfy the previous requirements to create an effective 15 minutes exposure time which mimics the expected exposure time for PLATO targets and strongly mitigates the contribution from the 5 minutes p-modes pulsations \citep{Chaplin(2019)}. Stacking the three consecutive spectra produces an effective visual bandpass SNR $\sim$450 or 15 cm/s of RV photonic precision. As an important note, the Sun-as-a-star on HARPS-N\footnote{Explained by the smaller telescope size and the usage of the integration sphere.} is equivalent to a magnitude $m_v=5.4$ on the 3.6m TNG telescope or to a magnitude $m_v=7.2$ on the 8.2m VLT telescope. Such magnitude represents the bright tail of the PLATO sample \citep{Nascimbeni(2026)}. 

In order to avoid obvious reverse engineering (such as doing the difference of two RV time series), three independent realisations of spectra selection was done for each of the planetary systems described in Sect.\ref{sec:planetary}. Each data set contains a different number of observations. We will call those three data sets hereafter A1050, B1300 and C800 related to the respective planetary system and number of data points each of those data set have (see Fig.\ref{fig:data sets} for the exact number). Note that the data set A, B and C represents a huge time investment with respectively 26, 33, and 21 equivalent nights of telescope time which is realistically the amount of time needed to be able measuring such HZ signals. The season length $S$ of each data set is also slightly different even though longer than half-a-year, with $S=225$, $S=236$ and $S=207$ days. This is roughly two months shorter than the season length expected from PLATO field from Paranal ($S\sim280$ days).

We displayed in Fig.\ref{FigGLS} the Generalized Lomb-Scargle (GLS) periodogram \citep{Zechmeister(2009)} in semi-amplitude of the three RV data sets without the planets injected and also indicate the ratio between the planetary semi-amplitude and the native semi-amplitude power at this period. Similarly, we fitted the planetary systems on the RVs without the planet injected in order to quantify the initial amount of power and degeneracy between the injected systems and the data sets. We reported the semi-amplitudes $K_{\text{raw}}$ in Table.\ref{tab:keplerians}, larger is the number, higher is the cross-talk between the native power spectrum and planetary signals.


The RV time-series of the three data sets with the planets injected are shown in Fig.\ref{fig:data sets}. We also displayed as a comparison below the MgII index\footnote{\url{https://lasp.colorado.edu/lisird/data/bremen_composite_mgii}} converted in plage filling factor using the time-series published in \citet{Cretignier(2024)}. We also displayed the solar TSI (see Sect.\ref{sec:tsi}) that is a mixture of faculae and sunspot filling factor.

\subsubsection{Processing of the data}
\label{sec:processing}

Injection of the planetary signals at the spectrum level may be challenging since to be exact, the planetary signal should be injected on spectra free of instrumental systematics. Indeed, an incorrect processing that would strongly underestimate the EPRV problem would be to simply shift the spectra by the Keplerian solution, injecting the planetary signals on the laboratory rest-frame systematics.

We used YARARA \citep{Cretignier(2021)} that is a post-processing pipeline analysing spectra time series to correct instrumental systematics at the spectrum level. As a short recall, YARARA starts by a preliminary standardization of the S1D order-merged spectra by fitting their continuum with the {\sc rassine} code \href{https://github.com/MichaelCretignier/Rassine_public}{\faGithub} \citep{Cretignier(2020b)}. During this stage, all the spectra are also interpolated linearly on a common wavelength grid in order to analyse and correct each wavelength bins time series independently. The code then creates a master spectrum, and fit on the residual spectra time series  multi-linear bases from proxies extracted by the pipeline itself. 

The YARARA correction of HARPS-N includes the corrections of cosmics rays, telluric lines, ghosts \citep{Dumusque(2015)}, interference pattern \citep[e.g][]{Thompson(2020)}, instrumental point spread function (PSF) distortions (see \citet{Stalport(2023)} or \citet{Klein(2024)}) such as those introduced by the change of the solar $v\sini$ along the year \citep{Collier(2019)} and stellar activity. Note that any correction can be canceled and added back on the spectra easily. This is mandatory since the stellar activity and change of the PSF should be on the spectra prior to their RV shift by the Keplerian solution. 

For each data set, we first ran YARARA on the spectra time series, remove the stellar activity and PSF correction, shift the spectra according to the Keplerian solutions, before adding back all the contaminations. We then ran a second time the pipeline on this new data set and offer to the participant two flavours of the data: a) the $YV0$ flavour that is uncorrected of anything and b) the $YVA$ flavour corrected of all the instrumental systematics except the stellar activity. Participants were free to choose between the two versions of the data they wanted to use. The data ($\sim$ 15 Gb) are freely accessible on the Google Drive\footnote{\url{https://drive.google.com/drive/u/1/folders/1Jbz-i21D81Ilr90w3AVSDzygctGuFnqo}} and Zenodo\footnote{\url{https://doi.org/10.5281/zenodo.20151946}}.

\subsubsection{HARPS-N solar S-index}
\label{subsubsec:sindex}

We provide the S-index time series extracted from the HARPS-N spectra, which is a standard diagnostic typically delivered alongside RVs for visible-band spectrographs. The S-index is directly related to the filling factor of active regions \citep{Cretignier(2024)}, although the line response remains relatively weak for solar-like activity, requiring careful spectral extraction. 

Indeed, the Sun-as-a-star observations obtained with HARPS-N are known to be affected by significant contamination from optical ghosts crossing the cores of the Ca II H\&K lines, from which this activity proxy is derived \citep{Dumusque(2021)}.

For the Sun, this contamination would in principle be of limited concern, as instrumental systematics are expected to remain largely stable over time. However, regular warm-up cycles of the HARPS-N detector modify the relative intensities of these ghosts reflection with the main physical orders, introducing not only an absolute contamination but also a time-variable component.

This contamination can be mitigated by estimating the flux between physical spectral orders on the CCD, as described in \citet{Dumusque(2021)}, or by using the YARARA framework, which includes dedicated corrections for such effects \citep{Cretignier(2021)}. We opted for this latter for simplicity.

In addition, the contamination is further reduced by applying the emission-profile extraction method presented in \citet{Cretignier(2024b)}. In contrast to the classical Mount Wilson S-index $S_{MW}$, which relies on a triangular bandpass integration, this approach directly fits on the spectra the emission profile of plage regions to the data \citep[see][]{Oranje(1983b),Cretignier(2024)}. This method provides improved uncertainties and is less sensitive to outliers compared to standard bandpass integration. In this framework, the plage emission profile used for the extraction can be interpreted as an optimally weighted integration of the signal. 

Note that, due to the spectrograph spectral efficiency response of HARPS-N, the SNR at 550 nm is equivalent to twice the value of the one in the continuum region surrounding the CaII H\&K lines at 420 nm.

\subsubsection{Solar TSI}
\label{sec:tsi}

We provided the total solar irradiance (TSI) as obtained from the LISIRD\footnote{\url{https://lasp.colorado.edu/lisird/data/nnl_tsi_P1D}} database. We did not inject the different transits in the TSI because this later is not necessarily representative of the precision expected from the PLATO lightcurves, adding an extra layer of complexity in the challenge and its possible interpretation that we wanted to avoid. Instead of that, as explained previously, we provided the exact phase and period of the transiting systems and mainly provided the TSI as a proxy that participants could use to fit their stellar activity model. 

\begin{figure}
	\centering
	\includegraphics[width=8.5cm]{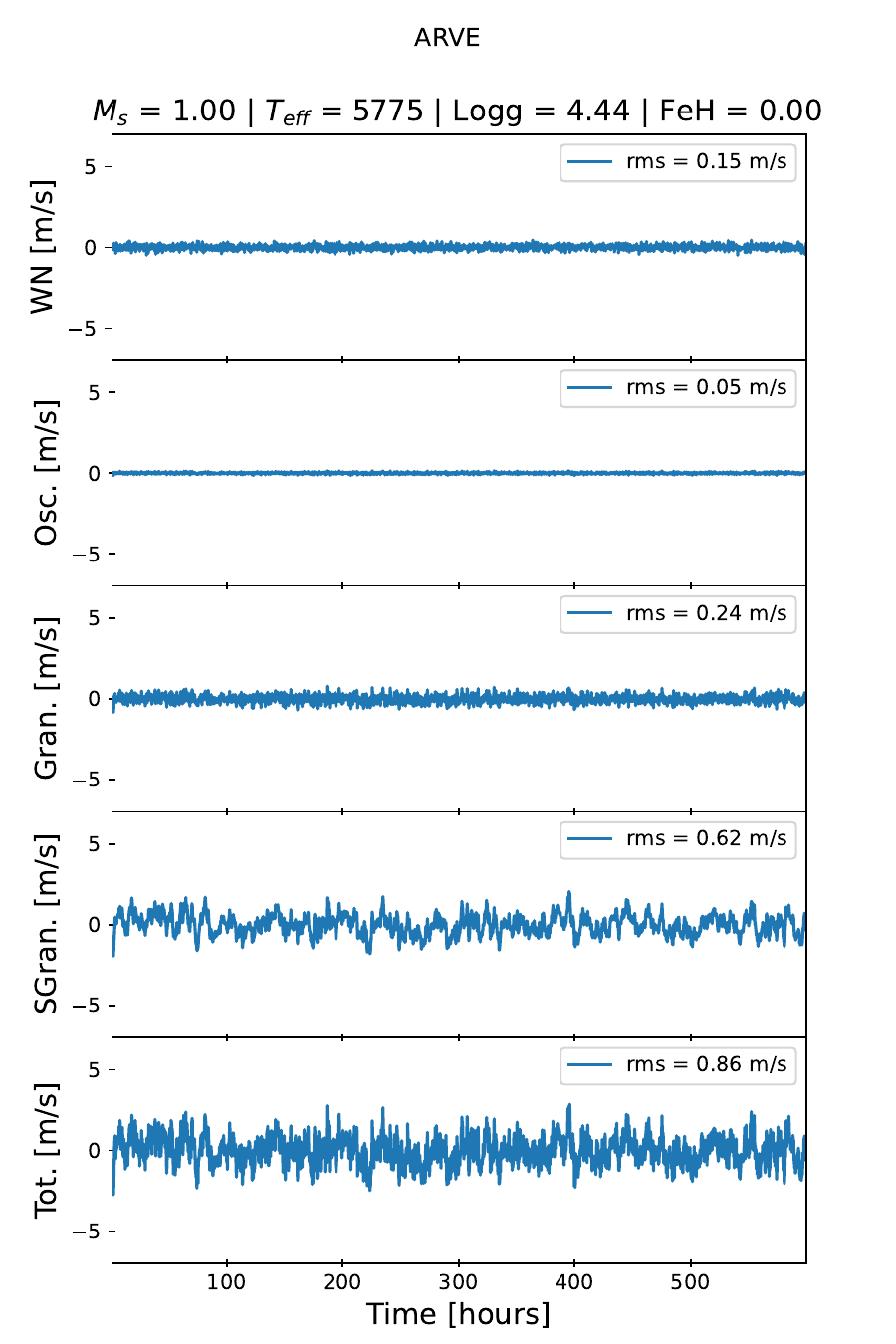}
	\caption{Representation of the RV budget expected from the main solar components in the PLATO RV challenge that uses the solar HARPS-N data set. A time serie of eight years including 15 minutes of integration time was generated using the ARVE code using the solar atmospheric parameters (only one rotational period is shown here). The contribution from each component is indicated in the legend with from top to bottom: the photon noise (15 cm/s), p-mode oscillation (5 cm/s), granulation (24 cm/s), supergranulation (62 cm/s) and their sum (86 cm/s).}
	\label{FigARVE}
 
\end{figure}

\subsubsection{ARVE solar photospheric components}
\label{sec:arve}

In order to benchmark the RV challenge and assess performance on a less demanding dataset, we generated a simplified red-noise model free of instrumental systematics and active regions: it only contains white noise, oscillations, granulation and supergranulation. For this purpose, we used the ARVE code\,\href{https://github.com/almoulla/arve}{\faGithub} \citep{Almoulla(2025)}, which simulates realistic time series from their power spectrum, as fitted on the HARPS-N solar dataset. 
The individual RV contributions are shown in Fig.~\ref{FigARVE} for a three weeks baseline. The model predicts a minimum total RV jitter of 86 cm/s, composed of 15 cm/s from photon noise, 5 cm/s from p-modes, 24 cm/s from granulation, and 62 cm/s from supergranulation. These contributions combine to produce the total RV budget of 86 cm/s similar to the budget found in \citet{Almoulla(2023)} and \citet{Meunier(2020c)}. We note that we did not include potential long-term magnetic cycle modulation of these components. Recent work suggests that supergranulation properties may vary with magnetic activity \citep{Osullivan(2025)}, but this effect is not accounted for in the present model. Using ARVE, we simulated the same 8-year RV time series with 15-minute integration times as the three data sets. This theoretical red-noise time series will be used (see next Sect.~\ref{sec:white}) for benchmarking purposes and represents an optimistic case scenario.

\subsection{Participants}
\label{sec:participants}

In total, eight participants participated in the challenge, seven provided a full set of results and one participant provided a partial result analysing only one (B1300) of the three systems. We compared the results of the candidates with a fictive processing represented by a benchmark participant. The output expected from the participants were the $K_i$ samplers from their guess Keplerian solution. We present here below the methodologies of all the participants to deliver those products. 

\subsubsection{Benchmark (exact Keplerian solution + ARVE)}
\label{sec:white}

This fictive participant is not using the YV0 or YVA data sets, but the ARVE RV budget described in Sect.\ref{sec:data set} that includes the same planetary signals on it. A Keplerian signal was added for each transiting planet with the period and phase specified by the challenge. Non-transiting planets were added if visible as a significant peak in the residuals periodograms (obtained after fitting the transiting planets amplitudes), which was the case for all the signals due to the lack of power for the ARVE time-series at long ($P>3$ days) periods. All the eccentricities were assumed null and uniform prior was used on all the remaining parameters. Note that we also investigated a benchmark with free eccentricities that provided results very similar and we therefore stuck to the null eccentricity model to a fair comparison with the participants. 

Under the assumption that no participant is able to deal with supergranulation (which is a question none of them have tried to answer so far), this participant represents the best performance that can be achieved in theory since the Keplerian model is close to the exact one and there is no long-term stellar activity and instrumental systematics. It will allow us to check if the amount of information in the time-series was sufficient or if solar photospheric components are sufficient to hinder the success of the challenge. All the MCMC analysis were done from the DACE\footnote{\url{https://dace.unige.ch/radialVelocities/}} implemented tool. Transit phases and periods were kept fixed for the transiting planets during the MCMC exploration.



\subsubsection{Linear regression of the S-index - (S. Aigrain)}\label{participant:sindex1}

The participant analysed both the YV0 and YVA data sets, assuming that all planets were on circular orbits, as the participants were informed of this \emph{a priori}. The analysis consisted of two steps: 1) detecting the non-transiting planets, and 2) jointly fitting for the planetary and stellar signals.  

In both steps, the stellar RV variations were assumed to be proportional to the variations of the S-index. 
In step 1, the RVs were first modelled as a linear combination of the S-index time-series, and sinusoidal terms corresponding to the known ephemerides of the transiting planets (ignoring mono-transits).
The coefficients of the resulting, linear basis model were fitted using {\sc scikit-learn.LinearRegression}\footnote{\url{https://scikit-learn.org}.} (version 1.2.2), and the best fit model subtracted from the RVs. The GLS periodogram of the residuals was then computed using {\sc astropy.timeseries.LombScargle}\footnote{\url{https://www.astropy.org/}.} (version 6.1.7), on a regularly spaced grid of frequencies from $\nu=0.0002$ to 0.9\,cycles/day ($P=1.1$--5000\,days), with {\tt nyquist\_factor}$=10$. If the highest peak in the periodogram had a False Alarm Probability ${\rm FAP} \le 10^{-5}$, the corresponding period was taken to correspond to an additional planet. A sine and a cosine term at this new period were added to the linear basis model, and the process was repeated until no further signals were found below the FAP threshold. 

The fitted coefficients of these additional terms were then converted to RV semi-amplitudes and $T_0$ values for the additional planets. 
For system 3, whose RV variations are dominated by non-transiting planets rather than activity signals or transiting planets, the initial fit was omitted and the initial GLS periodogram was computed directly on the YV0/YVA RVs. 

Step 2 consisted of a Bayesian modelling of the planetary and stellar signals using Hamiltonian Monte Carlo (HMC) to explore the joint posterior over all the parameters. The free parameters of the fit were thus: the systemic $RV_0$ offset, the semi-amplitudes $K_i$ of all the planets, the periods $P_i$ and times of inferior conjunction $T_{0,i}$ of the additional planets identified at step 1 only, the coefficient $S$ of the S-index term, and a jitter term $\sigma_{\rm jit}$ added in quadrature to the RV uncertainties. 
This participant used {\sc jaxoplanet}\footnote{\url{https://jax.exoplanet.codes/}.} (version 0.1.0) to evaluate the RV signatures of the planets.

The HMC sampling was implemented in {\sc numpyro}\footnote{\url{https://num.pyro.ai}.} (version 0.18.0) using the No U-Turn Sampler (NUTS). For the semi-amplitudes, half-normal priors with a mean of zero and a standard deviation of 50\,m/s were used. For the periods, the prior was a truncated log-normal with a standard deviation of 0.05 (in log), and bounds mid-way between the logarithms of the initial period guesses. This choice of prior allowed the model to explore a fairly wide range of periods for each planet without allowing the planets to "switch" order. A uniform prior between $0$ and $2\pi$ was used for the orbital phase of the additional planets, using {\sc numpyro-ext.distx.Angle}\footnote{\url{https://pypi.org/project/numpyro-ext/}.} (version 0.0.4) to sample this circular variable correctly. Wide normal priors (with a standard deviation of 10\,m/s) were used over $RV_0$ and $S$, and a log-uniform prior between 0.1 and 3\,m/s over $\sigma_{\rm jit}$. After a 1000-step burn-in phase, 2 chains of 10000 steps each were run, which was sufficient to ensure that the Gelman-Rubin statistic $R \le 1.01$ and the Effective Sample Size (ESS) was more than half of the number of samples for all parameters and all 3 systems. The resulting samples were provided as output along with various diagnostic plots (best-fit model and residuals, chains and corner plots).


\subsubsection{Linear regression of the S-index - (M. Damasso)}
\label{participant:sindex2}

The team analysed the YVA dataset only, assuming that all planets have circular orbits, which is a prior information provided to the participants. For all the targets, each test model included a linear regression with the $S_{\rm MW}$ index and the planetary Doppler signals, after verifying that a significant correlation exists between the RV and $S_{\rm MW}$ datasets. They also included a constant uncorrelated jitter term, which was added in quadrature to the nominal RV uncertainties. 

A frequency content analysis of the RV data was initially performed with \textsc{GLS} to identify potential signals due to non-transiting planets in the original datasets and pre-whitened residuals, having a FAP<0.01\% determined through a bootstrap analysis. These signals defined the maximum number of planets to be fitted for each simulated system through a Monte Carlo (MC) simulation.  
The MC modelling was performed by exploring the parameter space with the publicly available nested sampler and the Bayesian inference tool \textsc{MultiNest v3.10} (e.g., \citealt{Feroz2019}), through the \textsc{pyMultiNest} wrapper \citep{Buchner2014}. included 500 live points, a sampling efficiency of 0.3, and a Bayesian evidence tolerance of 0.5.  The package \textsc{radvel} \citep[v1.4.9][]{Fulton2018PASP..130d4504F} was used to model the RV Doppler planetary signals. 

Bayesian model comparison was performed to by calculating the difference $\Delta \ln\mathcal{Z}$ between the values of the Bayesian evidence $\ln\mathcal{Z}$ given by MultiNest for each model, assuming that the a-priori probability is the same for each model, and following the scale in \cite{feroz2011} to interpret the statistical significance $\Delta \ln\mathcal{Z}$. Values of $\Delta \ln\mathcal{Z}$ were calculated between pairs of models with a number of planets $n$ and $n+1$  
The team adopted uninformative priors for all the free parameters, except for the orbital period and epoch of inferior conjunction of the transiting planets, for which they adopted Gaussian prior distributions using the information provided by the organisers. In particular, we explored the semi-amplitudes of the planetary Doppler signals within intervals with lower limit zero and upper limits from 5 to 50 ms$^{\rm -1}$, depending on the $\chi$-square best-fit values calculated with \textsc{GLS} for each potential planetary signal. 

The team analysed the residual signal in RVs and found a significant stellar activity component that could be fit with a quasi-periodic GP. However, inclusion of the GP itself with the previously fit Keplerian systems did not significantly affected their results and decision was made ultimately to adopt in this work models with no GP term included in order to provide results from a framework not too sophisticated that can be regarded as a benchmark against which to compare those obtained using more complex models.

\subsubsection{Multi GPs - (L. Malavolta, D. Nadiello, G. Mantovan)}
\label{participant:multigp}

The participant analysed the YVA data set only and performed a joint modelling of photometric and spectroscopic data within a Bayesian framework using {\tt PyORBIT}\footnote{\url{https://pyorbit.readthedocs.io/}} \citep{Malavolta(2016),Malavolta(2016b),Malavolta(2018)}. The RV signal of each planet was modelled using a Keplerian function. For known transiting planets, the eccentricity was fixed to zero and the argument of pericenter to $90^\circ$. Arbitrary tight Gaussian priors were imposed on the orbital period and the time of inferior conjunction around the provided values.

For planets with unknown periods, a range from $10$ to $10^{2}$ days was explored in uniform log$_2$ space, with eccentricity allowed to vary up to 0.5. When the time of inferior conjunction was not available, the mean longitude was fitted instead, assuming a longitude of the ascending node equal to zero. The analysis included up to four additional non-transiting planets alongside the known transiting planet. For planets with known periods, the period and RV semi-amplitude were explored in linear space; otherwise, they were explored in log$_2$ space.

Stellar activity was mitigated using GP regression through the {\rm S+LEAF 2} package \citep{Delisle(2020),Delisle(2022)}. An Exponential-Sine Periodic (ESP) kernel (including up to four harmonics) was adopted as a computationally efficient approximation of the squared-exponential periodic kernel. For the spectroscopic dataset, a multivariate GP regression was implemented using full width at half maximum (FWHM), bisector-span (BIS), equivalent-width (EW), S-index, and \Halpha as activity indicators to constrain the underlying process.

For the photometric data, three scenarios were considered: exclusion of the photometry; GP regression using the same kernel and shared hyperparameters as the spectroscopic dataset but with an independent covariance matrix; and full inclusion within the multivariate GP framework, treated analogously to the spectroscopic activity indicators.

The first derivative of the GP was included only for the RV, BIS, and EW time series \citep[see e.g.][]{Mantovan(2024)}. To avoid degeneracies in the GP coefficients, the convective coefficient associated with the RV was constrained to be positive, while uninformative priors were adopted for all other parameters. Each dataset included independent offset and jitter terms. Additional tests considered both unidimensional GP models and configurations using only BIS and S-index as activity indicators.

The parameter space was explored using the affine-invariant MCMC ensemble sampler {\tt emcee} \citep{Foreman(2013)}, initialized from the results of the differential evolution optimizer {\tt pyDE} \href{https://github.com/hpparvi/PyDE}{\faGithub}. The choice of sampler was motivated by the need for controlled execution time. The optimisation phase was run for 50,000 generations with {\tt pyDE}, followed by 100,000 MCMC steps with {\tt emcee}, adopting a thinning factor of 100 and a burn-in of 80,000 steps. The number of chains was set dynamically to four times the dimensionality of the model.

Model selection was performed by comparing Bayesian Information Criterion (BIC) values derived from the RV-only likelihood, with the penalty term computed using only the parameters involved in the RV modelling. This approach was specifically implemented for the RV data challenge.

\subsubsection{CCFs SCALPELS -  (A. Collier-Cameron)}
\label{participant:scalpels}

This participant applied the {\sc scalpels} algorithm on the CCFs. Since some instrumental systematics can also induce variations in the CCF shape, both the YV0 and YVA realisations of the CCF time series are analysed.

The autocorrelation function (ACF) of a spectrum (or of a CCF) is invariant under translation. The {\sc scalpels} algorithm \citep{Collier(2021)} exploits this property to identify orthonormal modes of variation in the CCF shape that are insensitive to planetary Doppler shifts. The ACF, denoted $\boldsymbol{A}$, is computed for each CCF in the time series using a circular shift and cross-multiplication, with each row normalised by its mean. A singular value decomposition (SVD) of the resulting time series is then performed:
\begin{equation}
\boldsymbol{A} = \boldsymbol{U}\cdot\boldsymbol{S}\cdot\boldsymbol{V}^T.
\label{eq:svd}
\end{equation}

The rows of the right-singular matrix $\boldsymbol{V}^T$ represent orthonormal modes of variation in the ACF shape, while the columns of the left-singular matrix $\boldsymbol{U}$ provide the corresponding orthonormal time-domain coefficients. The diagonal matrix $\boldsymbol{S}$ contains eigenvalues ordered by decreasing contribution to the variance of $\boldsymbol{A}$. In practice, these eigenvalues decrease rapidly, such that most of the signal can be captured by a low-rank reconstruction using only the first few principal components, typically four.

The solar RV time series contains a small number of outliers, generally arising from anomalous distortions of the CCF that are poorly described by a low-rank model. These are identified and removed using an iterative scheme based on low-rank reconstruction and outlier rejection.

The shape-driven component of the RV signal is obtained by projecting the RV time series onto a basis defined by the leading columns of the matrix $\boldsymbol{U}$. These basis vectors are reordered according to their contribution to the Bayesian Information Criterion (BIC) of the RV fit. After this reordering, the first four columns of $\boldsymbol{U}$ provide a satisfactory reduced-rank representation of the shape-driven RV signal without significant overfitting.

The search for planetary signals is carried out using the {\sc kima} \href{https://github.com/kima-org/kima}{\faGithub} package \citep{Faria(2018)}. This framework simultaneously fits Keplerian orbital models together with linear combinations of the selected columns of $\boldsymbol{U}$, treated as activity indicators, using the diffusive nested sampling algorithm {\sc dnest4} \citep{Brewer(2015)}. Gaussian priors are adopted for the ephemerides of known transiting planets. When only the time of conjunction from a single transit is available, an uninformative prior is used for the orbital period, allowing the algorithm to identify any RV signal consistent with the observed transit time.

The trans-dimensional nature of the sampler allows the number of non-transiting planets to vary up to a maximum value $N_{\rm p,max}$. Bayesian odds ratios between models with different numbers of planets are derived from their relative occurrence in the posterior samples.

\subsubsection{CCFs LINSTAR - (J. Blanco-Pozo, M. Perger, I. Ribas)}
\label{participant:linstar}

The participants applied the \texttt{LINSTAR} (LINear Regression for STellar Activity Removal)
method, which is an adaptation of the CANSTAR method developed by \citet{BlancoPozo(2026)}
for this specific challenge. The CANSTAR activity correction relies on training a convolutional
attention network on StarSim \citep{Herrero(2016)} generated data; however, the lack of sufficiently
realistic synthetic models of the Sun at the time of the RV challenge led to an adaptation of
the method into a purely data-driven approach. The method operated through a two-step
procedure:

\begin{enumerate}
    \item Separation of line shifts from line-shape distortions in the CCF is achieved through the following decomposition:
    
\begin{equation}
\begin{split}
\boldsymbol{CCF}(\boldsymbol{v},t) =\;& c_0(t)\,\boldsymbol{I}(\boldsymbol{v}) + c_1(t)\,\boldsymbol{T}(\boldsymbol{v}) + l(t)\,\boldsymbol{T'}(\boldsymbol{v}) \\
&+ \sum^{N-1}_{n=2} g_n(t)\,\boldsymbol{G_n}(\boldsymbol{v}) + \boldsymbol{E}(\boldsymbol{v},t),
\end{split}
\label{eq:final}
\end{equation}
    
where $\boldsymbol{I}$ is the unity function, $\boldsymbol{T}$ is the time-averaged template CCF, $\boldsymbol{T'} = d\boldsymbol{T}/d\boldsymbol{v}$ is its first derivative (capturing the line shift $l(t)$), and $\boldsymbol{E}$ is the residual error. The functions $\boldsymbol{G_n}$ form a basis describing line-shape distortions. A Gram--Schmidt orthogonalisation is applied to ensure that $\boldsymbol{I}(\boldsymbol{v})$, $\boldsymbol{T}(\boldsymbol{v})$, and $\boldsymbol{G_n}$ are strictly orthogonal to $\boldsymbol{T'}$, enabling a clear separation between Doppler shifts ($l$) and pure line-shape distortions ($[c_0, c_1, g_2, \dots, g_N]$).

    \item A joint fit of Keplerian orbits and a linear combination of the selected line-shape distortion coefficients ($[c_0, c_1, g_2, \dots, g_N]$) using the dynamic nested sampler \texttt{dynesty} \citep{speagle2020dynesty}. 
\end{enumerate}

    For FGK stars, where $\boldsymbol{T}$ is approximately Gaussian—as in the present study—a Hermite basis is adopted. This basis is constructed from the $n$-th derivatives of a Gaussian function fitted to $\boldsymbol{T}$, starting from $n=2$, since the $n=0$ and $n=1$ components correspond to $\boldsymbol{T}$ and $\boldsymbol{T'}$. As shown in \citet{BlancoPozo(2026)}, these coefficients naturally map onto classical indicators ($c_1$ to contrast, $l$ to RV, $g_2$ to FWHM, $g_3$ to BIS), while extending to higher-order shape variability ($n \geq 4$), which captures both symmetric and asymmetric distortions.

Following a SCALPELS-like strategy \citep{cameron2021scalpels}, the optimal number of distortion coefficients is determined iteratively by minimizing the RV root-mean-square (RMS) while controlling model complexity through the Bayesian Information Criterion (BIC), thereby avoiding overfitting. Giant planets dominating the time-series variability are pre-whitened via a preliminary least-squares fit prior to this selection.
    
Once the optimal number of distortion coefficients is established, Keplerian signals are iteratively fitted to the data. For transiting planets, informative priors are applied on periods and phases. For non-transiting planets, uninformative priors are adopted for the orbital phase and semi-amplitude, while period priors are restricted to broad intervals around the dominant peaks in the periodogram to prevent overlap between signals in multi-planet fits.
    
The dynamic nested sampling is initialized with 500 live points, using a random-walk proposal to explore the parameter space. The sampler computes the log-Bayesian evidence ($\Delta \ln \mathcal{Z}$) by integrating the likelihood over the prior volume, progressively focusing on regions of higher likelihood. The log-likelihood values of the sampled points are stored at each iteration to compute importance weights, which are then used to derive the posterior distributions of the model parameters.
    
Model complexity is increased iteratively, evaluating both $\Delta \ln \mathcal{Z}$ and the residuals after subtracting the best-fitting model. The addition of further components is halted when no significant signals remain in the residual GLS periodogram (FAP $> 1\%$), or when the increase in model evidence no longer justifies the added complexity \citep[$\Delta \ln \mathcal{Z} \le 5$;][]{trotta2008bayes}.



\subsubsection{SHELL CNN - (Y. Zhao)}
\label{participant:shell}

The shell-CNN is a convolutional neural network (CNN)-based method designed to model stellar activity signals at the spectral level, thereby improving the detection of Earth-like planets \citep{Zhao(2024b)}. The approach relies on the distinction that planetary signals induce pure Doppler shifts in stellar spectra, while stellar activity produces variations in spectral line profiles, such as changes in asymmetry and depth.

For each dataset, the 1D solar spectra are projected into the spectral shell domain $(f_{0}, \frac{d f_{0}}{d v})$. A solar line mask derived from \citet{Cretignier(2020a)} is used to select strong, unblended spectral lines that are free from telluric contamination. 

Next, for each dataset, we trained a convolutional neural network from scratch to model the stellar activity induced RVs, full width at half maximum (FWHM), and bisector span (BIS) using a 10-fold cross-validation scheme. Specifically, the spectral shell data were randomly divided into 10 groups. The model was then trained over 10 iterations, where in each iteration one group was designated as the test set while the remaining nine groups were used for training. The final fitting results were obtained by concatenating the predictions on the test sets from all iterations.

The trained neural network is used to estimate and subtract the stellar activity–induced RV component from the total RVs, which are measured using a classical CCF applied to the solar spectra. The planetary signal is assumed to follow a circular orbit. A periodogram of the resulting RV residuals is then computed, and a numerically estimated FAP is derived, adopting a detection threshold of 0.1\%. A signal is considered detected if the corresponding peak exceeds this threshold. In such cases, a sinusoidal model is fitted to recover the amplitude and phase of the injected planet. Posterior distributions of the best-fit Keplerian parameters are subsequently obtained using the \texttt{UltraNest} package \citep{Buchner(2021)}.

For datasets A1050 and C800, the RV time series extracted from the solar spectra exhibit strong planetary signals. These dominant signals are first modelled with a single-planet fit and removed prior to training the neural network on the residuals, in order to isolate the stellar activity component.

\subsubsection{YARARA V2 - (M. Cretignier)}
\label{participant:yv2}

YARARA V2, presented in \citet{Cretignier(2023)}, is an extension of YARARA V1 \citep{Cretignier(2021)}. The first version primarily consists of the corrections described in Sect.~\ref{sec:processing}, applied directly in the wavelength domain at the spectral level. Once the spectra are corrected, line-by-line (LBL) RVs are extracted following \citet{Dumusque(2018)}. Time-domain proxies are then derived by projections onto compact spaces.

The first compact space, called the shell, corresponds to a change in the variables of the spectrum $f(\lambda)$ from the wavelength $\lambda$ to the flux gradient, $f\left(\frac{df}{d\lambda}\right)$. This transformation, introduced in \citet{Cretignier(2022)}, provides two main advantages. First, it preserves information on the normalised flux, which is an important diagnostic of stellar activity as an indirect proxy for photospheric depth \citep{Cretignier(2020a),Moulla(2022),Almoulla(2024)}. Second, in this representation, a pure Doppler shift has a well-defined form, referred to as the \emph{Doppler shell}, $DS(f, df/d\lambda)$, derived from the first-order approximation of the Doppler shift. A principal component analysis (PCA) is performed in this space to obtain components $PC(f, df/d\lambda)$. The number of components is selected using a modified $p$-leave-out cross-validation scheme (see Appendix of \citet{Cretignier(2022)}). A Gram–Schmidt orthogonalisation is then applied between these components and the \emph{Doppler shell} before fitting them to the data. The resulting coefficients form time series that have been shown to correlate with RV variations induced by activity or instrumental systematics.

The second compact space is constructed from the LBL RVs. Under the assumption that a planetary signal corresponds to a pure Doppler shift affecting all spectral lines identically, the variance across LBL RVs is free of planetary contributions \citep{Cretignier(2023),Ould-Elhkim(2023)}. Consequently, a PCA performed on LBL RVs, which operates on mean-centered data and captures differential variations between spectral lines, should not contain planetary signals. This again defines a time-domain basis that can be fitted to the data.

YARARA V2 consists of fitting these two time-domain bases to the LBL RVs. Transiting planetary signals are explicitly included in the model to prevent absorption of their amplitudes and to enable the search for additional signals in the residuals. A sinusoidal signal search is performed using an iterative GLS periodogram with a 1\% FAP threshold, limited to a maximum of six signals and orbital period range $P$ between 3 to 5000 days. Once the Keplerian model and the amplitudes of the basis vectors are fixed, the residual RVs are analysed using an MCMC framework, assuming circular orbits and fitting only for the amplitudes of the transiting signals with the DACE plateform as for the benchmark in Sect.\ref{sec:white}.

\subsubsection{1D GP - (E.C. Gonçalves)}
\label{participant:tm}

The participant only analysed the data set B1300 by lack of time. They used the {\sc kima} \href{https://github.com/kima-org/kima}{\faGithub} code \citep{Faria(2018)} to model the provided RVs with a quasi-periodic kernel. The model included 4 Keplerians for the transiting planets and a quadratic polynomial for the mean trend.
The Keplerians had variable period $P$, $K$ semi-amplitude and $T_c$ time, the prior for the periods was gaussian with mean being the given period value with a std of 0.1 (the mono-transiting case was set with a uniform prior from 100 days to 5 times the time span of the time series), the $K$ semi-amplitude had a uniform prior from 0 to 5 times the variation of the RVs and the prior for the Tc were gaussian with the mean being the given value and an std of 0.001. (eccentricities were set to 0 and argument of periastron to pi/2).

A standard quasi-periodic kernel \citep{Rasmussen(2006)} was used for the GP with the following priors for the hyperparameters $\eta_1$ set to default by {\sc kima} (uniform from 0 to the variation of the data max-min), $\eta_2$ set with a uniform prior between 15 and 600 days, $\eta_3$ set with a uniform prior between 15 and 50, $\eta_4$ was left as {\sc kima}'s default which is a uniform prior between 0.2 and 5.

The model was run using diffusive nested sampling \citep{Brewer(2016)} for 1 000 000 steps with 16 threads.

\section{Results of the challenge}
\label{sec:results}

In this section, we present the analysis of the three data sets processed by the participants. We start by defining a score metric (Sect.\ref{sec:secmetric}) that can be used in the context of the PLATO mission in order to compare participants (Sect.\ref{sec:result_individual} and Sect.\ref{sec:ranking}). Later, the results of all the participants will be put together in order to get broader conclusions for the RV field (Sect.\ref{sec:rvfield}).  

\subsection{Defining a score metric}
\label{sec:secmetric}

The output asked to the participants were the posterior samples 
of the Keplerian fit on the three data sets. We will define a \emph{false detection} as any fitted signal with a period farther than 10\% of the injected orbital periods $P_{\text{inj}}$. 
On the contrary, a \emph{missed signal} will be an injected planetary signal for which no signal were closer than those 10\% in period.

We then developed a metric using the semi-amplitude samples $K_i$ to evaluate the methods aiming to closely match the PLATO requirements written in Sect.\ref{sec:requirements}. We note that there is an ambiguity in the requirement, it is unclear whether the 10\% figure pertains to the precision or accuracy of the measurement. On the one hand, the accuracy cannot be checked in practice since the ground truth is never known. On the other hand, precise measurements can be biased, such that the confidence in the estimated value is overestimated.  The present challenge provides a good benchmark to link the accuracy, that can be computed here since the ground truth is known, to the precision, computable also for real PLATO follow-up data.


In this work, we will define the \acc\ by $\Delta K /K$ as the relative difference between the injected ground truth $K_{\text{inj}}$ and the median of the posterior on the samples $K_i$:
\begin{equation}
    \Delta K /K= \frac{K_{\text{inj}}-med(K_i)}{K_{\text{inj}}}
\end{equation}
Note that our accuracy definition is signed, but we also use an unsigned one. It will be clear when signed versus unsigned accuracy was used.

\begin{figure*}
	\centering
	\includegraphics[width=18cm]{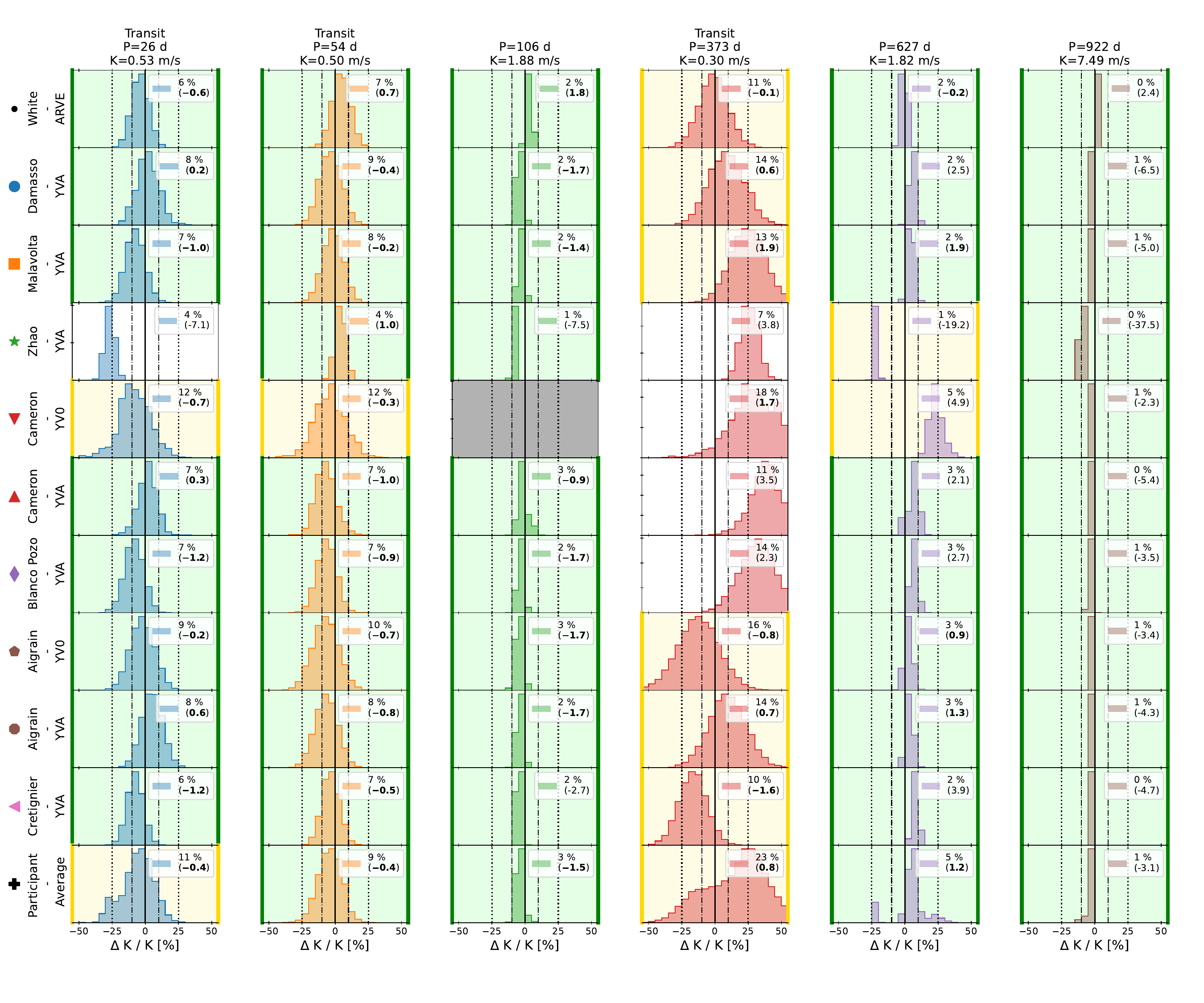}
	\caption{Representation of the $K_i$ semi-amplitude posterior for the six injected planets of the A1050 data set for each participant. Black vertical line was the ground truth $K_{\text{inj}}$, while 10\% and 25\% levels are indicated by the vertical dashed-dotted and dotted lines respectively. All the panels with a score better than 25\% are displayed in yellow box, while all the scores better than 10\% are in green. Planets missed are showed with a black frame. Each panel also indicates the \acc\ and the number of sigma $N_\sigma$. If $N_\sigma$ is smaller than 2, the value is highlighted in bold. For most participants all the planets are constrained down to 10\%. The most difficult signal of the A1050 data set is the HZ transiting signal at 373 days for which no participant get a 10\% score and the mass tend to be overestimated. We do not see any peculiar issue for the transiting planet closed to the solar rotation period.}
	\label{fig:results1}
\end{figure*}

On the other hand, we defined the \prec\ $\sigma_k$ as the sigma width of the posterior, measured by the median absolute deviation (MAD) of the samples, normalised by the ground truth injected value:
\begin{equation}
    \sigma_K = \frac{1.48 \times MAD(K_i)}{K_{\text{inj}}}\, ,
\end{equation}
where the factor 1.48 is just a conversion factor from MAD to an usual  standard deviation for a Gaussian distribution. 

While it is trivial that neither \acc\, or \prec\, alone are sufficient to assess the score of a participant, it is certain that a method for which both the \acc\, and \prec\, are good should be considered as successful in general. We therefore defined the \score\ $\varepsilon_k$, of the challenge as the worst value between the two:

\begin{equation}
    \varepsilon_K = max\left(|\Delta K / K|,\sigma_K\right)
\end{equation}

Note that with those definitions, the usual "number of sigma" $N_\sigma$ is just given by the ratio of the \acc \, over the \prec: 

\begin{equation}
N_\sigma = \frac{|\Delta K / K|}{\sigma_K}
\end{equation}

If in practice this \score\ $\varepsilon_k$ definition is an attempt to obtain a one-dimensional metric to rank the participants, as described later, this ruler is not perfect and we therefore need to always present the pair ($\Delta K / K$, $\sigma_k$) or their alternative representation ($\varepsilon_k$, $N_\sigma$) to get a global view of a method. 

\subsection{Results of the participants}
\label{sec:result_individual}

In Fig.~\ref{fig:results1}, we display the $K_i$ posteriors distributions for the data set A1050 for all the correctly identified signals for each participant. The question of \emph{false positives} will be discussed later. For later comparison (see Sect.~\ref{sec:rvfield}), we also created the average posterior from all the participants. Each participant is represented by a unique color and marker shape displayed on the left and preserved along the paper.

Until the end of the paper, we defined two arbitrary levels as relevant for the PLATO mission. The first level is the 10\% level that matches the PLATO requirements, the second level is 25\% \score \, that is the score level needs to achieve accurate planetary atmosphere modelling \citep{Seager(2010),Crass(2021)}. We display in the Fig.~\ref{fig:results1}, the results satisfying 10\% by a green box, and the ones satisfying 25\% by a yellow box.  

As visible in the figure, for all the signals larger than 50 cm/s in amplitudes the 10\% \score\ $\varepsilon_K$ is achieved by most participants. However, such a score is not achieved for the HZ transit planets that reach at best the 25\% \score. On a positive note, the transiting planet close to the rotational period of the Sun is recovered at 10\% by most participants. 

\begin{figure*}
	\centering
	\includegraphics[width=18cm]{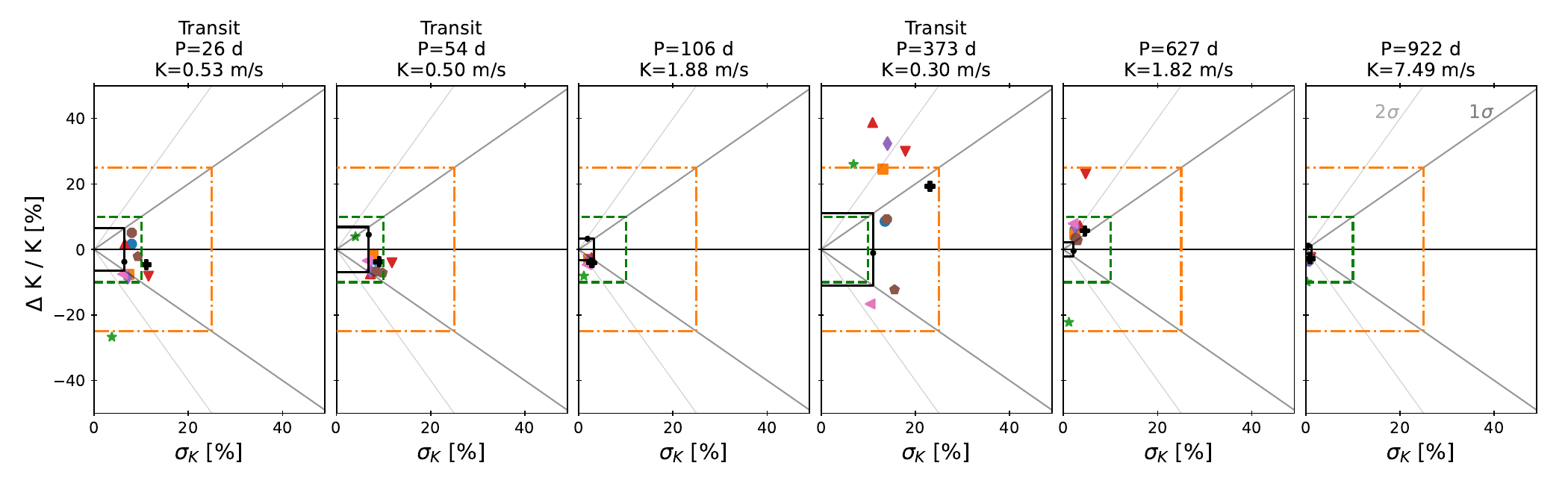}
	\caption{Representation of the pair \acc\, and \prec\, for each injected signal correctly detected by the participants for the data set A1050. Participants are represented with the same symbols as in Fig.\ref{fig:results1}. The 10\% PLATO requirement is indicated with a green dashed square, the 25\% score is shown with the orange dashed-dotted square. Any dot inside the square satisfy the score requirement. The black solid box represents the score of the benchmark (see Sect.\ref{sec:white}). The 1 and 2-$\sigma$ are shown by the cone plain lines. }
	\label{fig:playground1}
\end{figure*}

\begin{figure*}
	\centering
	\includegraphics[width=18cm]{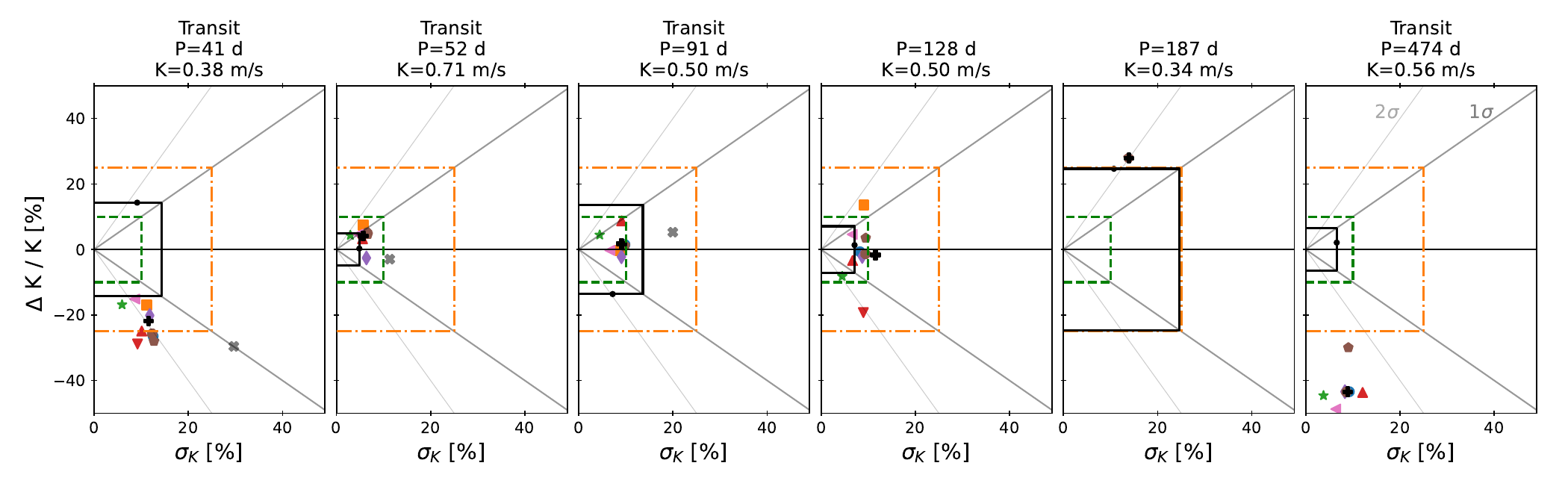}
	\caption{Same as Fig.\ref{fig:playground1} for the data set B1300. All the participants underestimated the mass of the inner most and outermost planets. If the outermost period was not constrained since the case of a mono-transit. The underestimation of the inner one is more difficult to understand. No participant except one, detected the non-transiting 187-day planet.}
	\label{fig:playground2}
\end{figure*}

\begin{figure*}
	\centering
	\includegraphics[width=15cm]{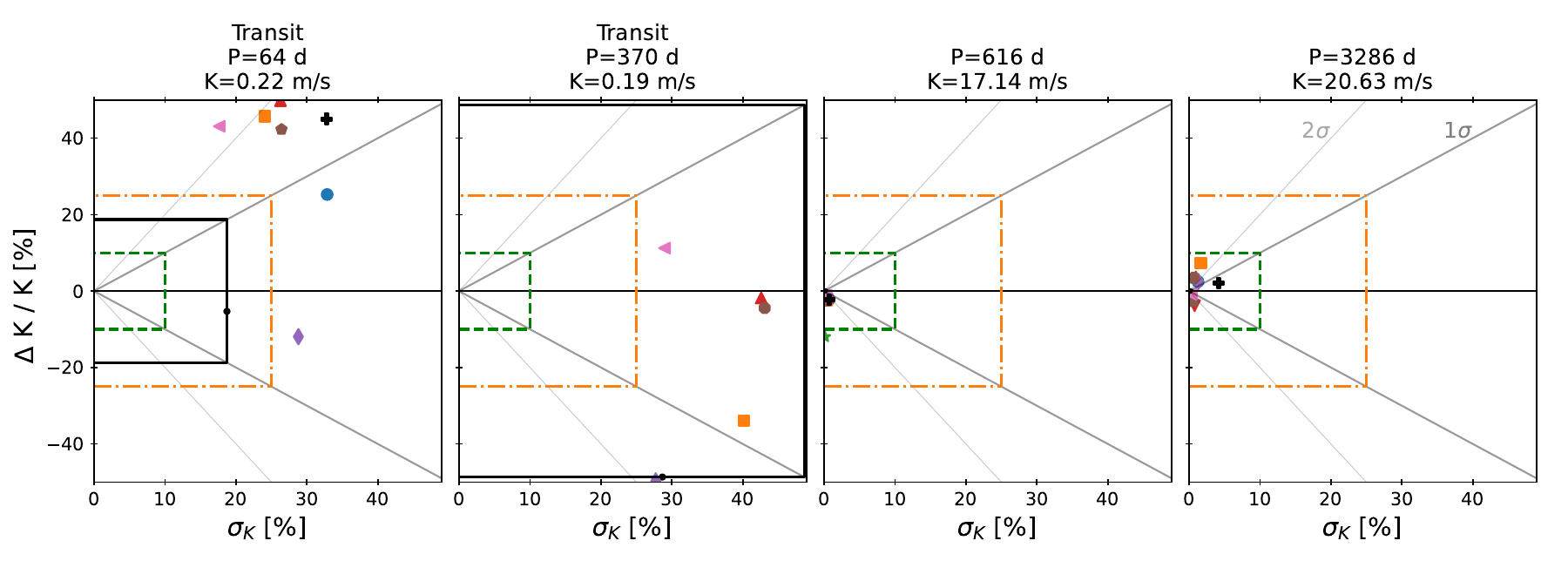}
	\caption{Same as Fig.\ref{fig:playground1} for the data set C800. With among the smallest amplitudes, no participants manages to reach 25\% precision on the transiting terrestrial planets. }
	\label{fig:playground3}
\end{figure*}

In order to better represent the \acc\ and \prec, we extracted the value for each participant and represented all of them in Fig.\ref{fig:playground1}. The same was done for the data set B1300 and C800, respectively in Fig.\ref{fig:playground2} and Fig.\ref{fig:playground3}. The 10\% and 25\% level are now represented by colored boxes. The score reached by the benchmark is also shown as a black box for comparison.

Such a visualization allow a rapid diagnostic of the easy and challenging planets. For all the data sets, a \score\, of 10\% was never achieved for any of the HZ planets and only the HZ planets of the system A1050 reach the 25\% \score. This confirms the results obtained by \citet{Meunier(2019e)} based on simulations that achieving such a precision level on the mass, even in the case of a transiting planet, is extremely challenging today. Our result even goes one step further, first by using this time real data, secondly by offering a broader range of methods tested. We first discuss below the results obtain on each system individually ordered by number of observations. 

\subsubsection{Results on C800}
\label{sec:result_c800}

This system was certainly the most difficult with semi-amplitudes as small as 20 cm/s. As visible in Fig.\ref{fig:playground3}, none of the participants reached the 25\% level for the transiting planets. On the other hand, most participants correctly recovered the amplitude of the giants, despite the existing cross-talk between the 11-year gaseous planet and the solar magnetic cycle. The phase of the 2 $M_{Jup}$ signal was slightly anti-correlated with the solar magnetic cycle ($\mathcal{R}=-0.61$) indicating some risk of interaction between the signals that was correctly accounted by the participants.

The performance on the transiting HZ planet is relatively poor, with a best score of only $\sim 30$\%. However, even the benchmark analysis does not exceed $\sim 45$\% for this signal\footnote{Some participants may occasionally outperform the benchmark, as residual red noise—particularly near a one-year period—can spuriously enhance or suppress the recovered signals depending on its phase relative to the Keplerian signal.}, indicating an intrinsic difficulty in modelling the data given the available information.

This low performance is likely driven by the proximity of the signal to a one-year period, which results in poor phase coverage. This effect is exacerbated by the observational window of C800, which spans only about 7 months, leaving approximately $\sim 40$\% of the orbital phase unobserved.

A more detailed analysis shows that one of the two quadratures is entirely missing. These phases are particularly informative for constraining the semi-amplitude $K$, and their absence significantly degrades the ability to recover the signal accurately.  

The only bad results are obtained for the 64-day transiting signal for which the average participant score is around 40\% while the benchmark reaches 20\% with a global overestimation of the mass. This likely shows that with realistic data, even 800 measurements (21 nights) are not enough to recover such small amplitudes signals even when phases and periods are known.

\subsubsection{Results on A1050}
\label{sec:result_a1050}

The results obtained on this data set are generally speaking the best as displayed in Fig.\ref{fig:playground1}. The levels of 25\% (and even 10\%) is reached several times by the participants. This is notably the case for the 26-day period planet close to the solar rotation period. While it should be reminded that most of our HARPS-N dataset is observing a quiet Sun from among the quietest cycles history \citep{Jiang(2015),Pesnell(2016)}, this seems to confirm than, nonetheless, 50 cm/s can be relatively well constrained in RVs with an HARPS-like stability precision instrument. 

As for C800 (see previous Sect.~\ref{sec:result_c800}), the worst result is obtained for the 373-day transiting HZ signal with an average score around 25\%, while the benchmark grazes the 10\% limit. Opposite to the signal in C800, this signal seems better constrained by both the benchmark and the participants. While the presence of 250 new observations (5 extra nights of telescope time) in A1050 compared to C800 should certainly help, it is doubtful that such an improvement can only be obtained by more data and part of the explanation can also be found in the phase coverage of the signal. Indeed, opposite to the signal in C800, the 8 months season length is now placing the seasonal gap no more on the quadrature of the signal but on the rising section of the phase. The presence of both quadratures certainly helping in the determination of the $K$ semi-amplitude.  

Despite the good average score value, this 373-day signal is also the one for which the score of the participants is most fluctuant, with score spanning between 15\% to 40\%. 

\subsubsection{Results on B1300}
\label{sec:result_b1300}

This data set is the least successful. All the participants performed significantly worse than the benchmark, in particular for the mono-transit HZ signal at 474 days, the inner most planet at 41 days, and the almost non-detection by all of the 187 days signal. This system definitely contained more complexity than expected.

This result was not anticipated since this data set is the only one for which all the signals are correctly covered in phase, contains the largest number of observations ($N=1300$ or 33 nights of telescope time) and has the longest season (8 months), so understanding why the signals were not recovered is a key point. 

It should be recalled that this HZ planet was the unique mono-transit test performed, meaning that no tight prior on the period was used by the participants, even if this signal was also the most massive transiting HZ of the challenge (7\Mearth) and correctly identified by most participants. As an interesting observation, the partial submission participant has attributed the mono-transit to the wrong planets (the 128-d signal) which can be explained by the lower boundary of the period prior spanning down to 100 days. 

Furthermore, in this system, almost all the participants used a wrong planetary system model by missing the injected signal at 184 days of 34 cm/s and included a false detection at 630 days. It is very likely that such a wrong planetary model combined by the broad period prior is at the origin of the poor performance achieved on this data set.

\begin{figure*}
	\centering
	\includegraphics[width=18.5cm]{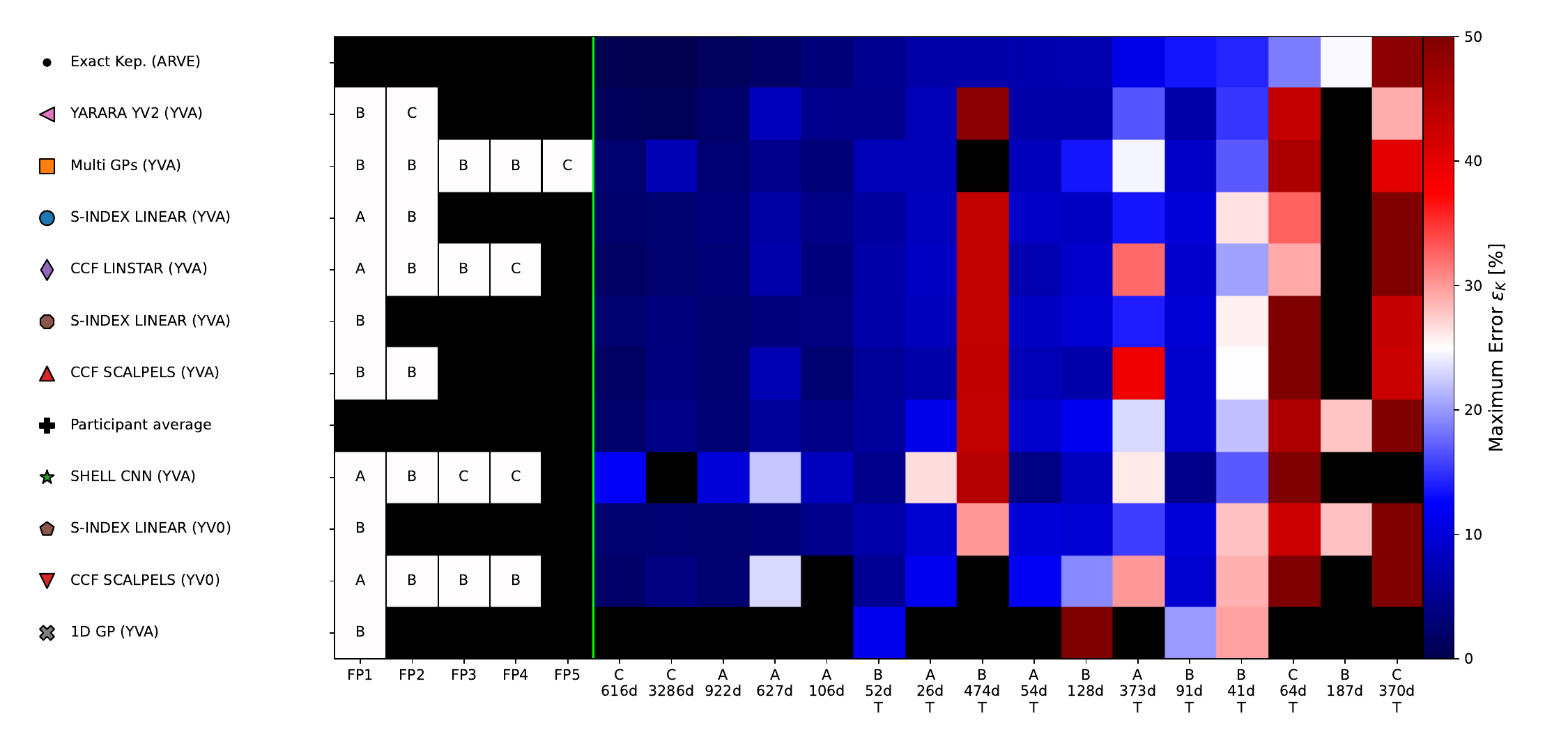}
	\caption{Representation of the \score\ $\varepsilon_K$ obtained by all the participants on each injected planetary signal. Each participant is represented by a row and its unique assigned color/marker combination. The participants are ordered by their average \score \, on the terrestrial signal category, but this ordering does not represent a ranking (see plain text Sect.\ref{sec:ranking}). The exact planetary signal is indicated in the bottom axis label by its system and orbital period (as well as 'T' transit indication). The planetary signals were ordered by the \score\ obtained by the benchmark (first row). The false positive detections are indicated on the left part of the vertical green line. The scale of the colorbar was set such as the 25\% level is white, better score will be blue, worse red. A missed signal (true negative) is represented by a black box.}
	\label{fig:participant_score}
\end{figure*}

By analysing the system architecture, we can notice a few peculiarities. The first is that three of the signals are close to the 1-year harmonics at 183, 122 and 91 days respectively. While not enough to explain the discrepancy (in particular since no eccentric signals were involved in this challenge), we can imagine that some 1-year systematics interact in some unpredictable way with the planetary signals. Such cross-talk were already mentioned for the previous data set except than this cross-talk is now occurring on a non-transiting planet.

We can question if the reduction by YARARA producing the YVA data set is not affecting those 1-year harmonics signals. This observation is partially motivated by the fact that the only participant that detected the 187-day signal used YV0 and not YVA, but we note that other participants using YV0 also missed this signal though. Furthermore, this explanation would not be coherent with the fact that no significant difference was observed between YV0 and YVA for the system A and C that were both containing 1-year planets.

\subsubsection{Global comments}
\label{sec:global}

There are several culprits that may explain why no method is today yet able to reach the required precision. 
For some of the transiting signals, the difficulty may arise due to their proximity with 1-year that is an expected very difficult case scenario due to aliasing, poor phase coverage of the signal with season gaps and systematics all producing strong interaction around the 1-year period. Such an issue mainly arise for the HZ of solar-type stars, which can be mitigated when observing cooler K-dwarf for which the HZ moves inwards (see Sect.\ref{sec:kdwarf}).

Part of the long-period difficulty may be due to imperfect stellar activity correction, even if we tested a rotational period transiting planet scenario. Similarly, supergranulation that is expected to produce part of the jitter observed in the HARPS-N solar data set \citep{Almoulla(2023)} may not leak power on such long timescale except if the properties of the supergranulation may also evolve on long timescale \citep{Osullivan(2025)}. 

While other exotic stellar activity phenomena may be invoked (such as meridional flows, see e.g \citet{Meunier(2020)}), the most likely culprit is the instrument stability itself, as well as granulation and supergranulation. It should be noted that while short-term stability of HARPS-N is known to be better than 50 cm/s \citep{Dumusque(2026)}, the long-term stability of instruments is very difficult to assess. The result of this challenge may indicate that the long-term stability of HARPS-like instruments is not sufficient to achieve the PLATO requirement for such small signals and more stable spectrograph (such as ESPRESSO) have certainly to be considered or further improvements should be achieved in this domain.

Given the result of the participants, one could infer the level of white noise which would yield the same $1\sigma$ uncertainties on the orbital elements (see Table.\ref{tab:keplerians2}). To compute this equivalent white noise, we use the analytical formula giving covariance of a least square fit. It has been tested that for the systems at hand, with over 1000 points, this analytical formula is consistent with a full MCMC computation of uncertainties. 
For system $A$ and $B$ and the time series generated with RV, we find that the corresponding level of white noise is 80 cm/s, in accordance with the simulation, and the YARARA YV2 method yields the same equivalent white noise value, while maintaining  a comparable accuracy. In both cases, the amplitude is consistant with the combined effect of p-modes, granulation and super-granulation, the latter one being dominating. To go further, it is therefore extremely important to focus on these signals.

\begin{figure}
	\centering
	\includegraphics[width=8cm]{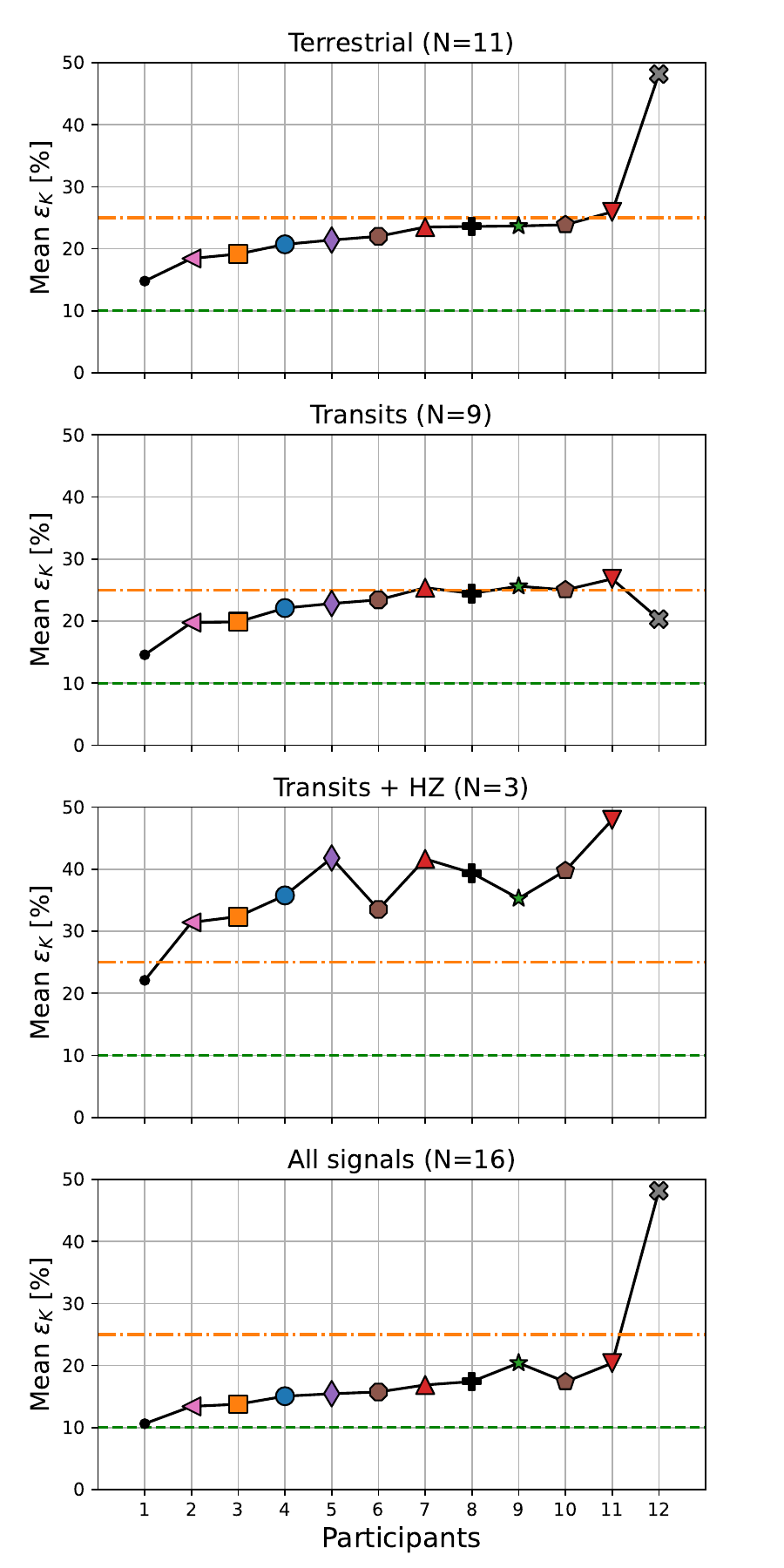}
	\caption{Average \score\ $\varepsilon_k$ for the 16 injected planets. Four different subsets were analysed (from top to bottom panels): terrestrial planets, the transiting planets, the habitable-zone (HZ) transiting planets and the full sample of planets. Each participant is represented by the same marker as in Fig.\ref{fig:participant_score}. Participants names are however not displayed since this metric alone is not sufficient to allow for a ranking (see plain text).  
    The two arbitrary levels defined in this challenge are also displayed as horizontal lines at 10\% (green dashed line) and 25\% (orange dashed-dotted line).}
	\label{fig:ranking_score}
\end{figure}

\begin{figure}
	\centering
	\includegraphics[width=8cm]{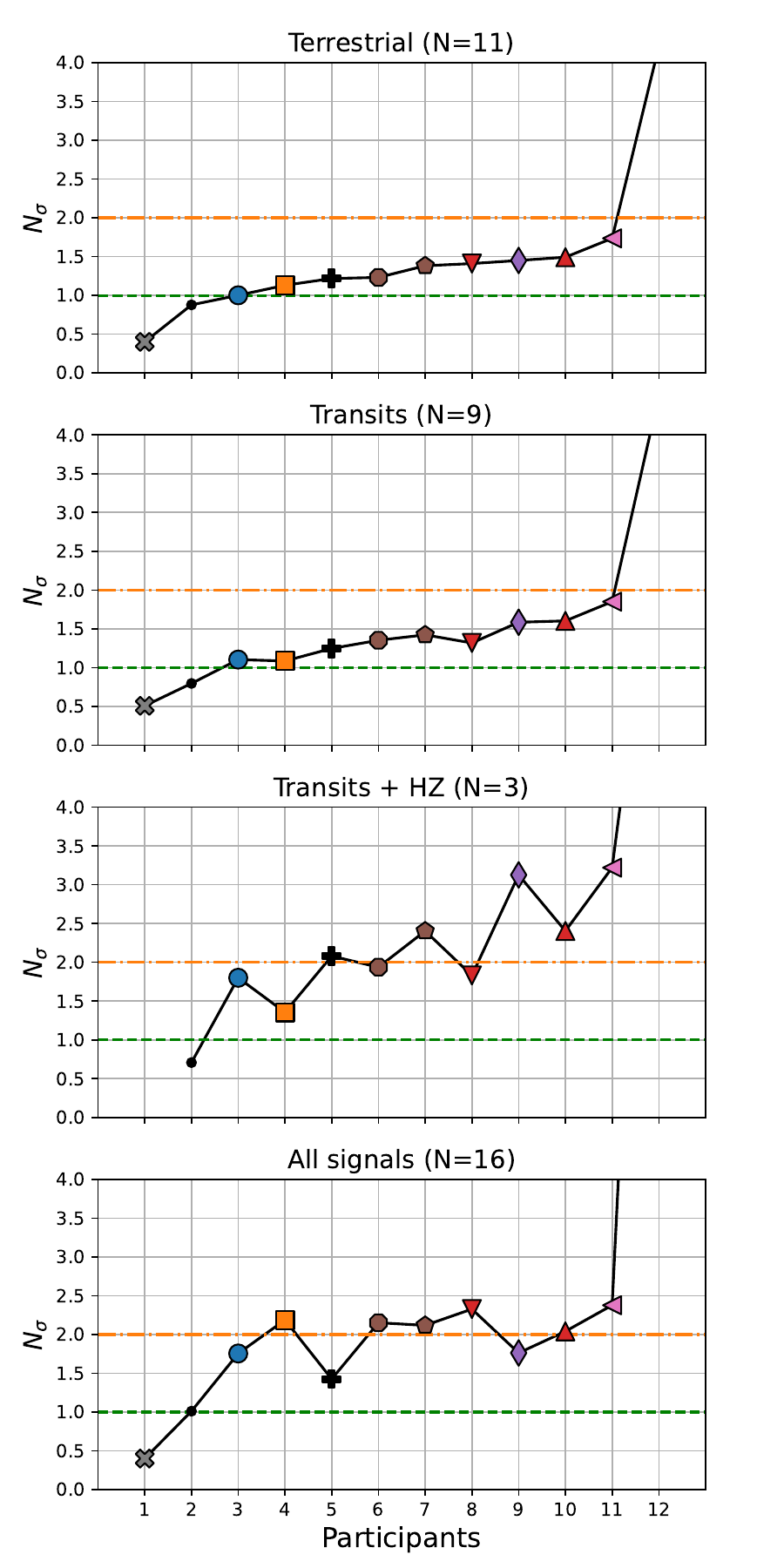}
	\caption{Same as Fig.\ref{fig:ranking_score} for the number of sigma $N_\sigma$, instead of \score\ $\varepsilon_k$. The 10\% and 25\% horizontal levels were replaced by 1 (dotted-dashed line) and 2 (dotted line) sigma level that are classical levels. A value above $2\sigma$ indicate an underestimation of the uncertainties and overestimation of the precision. A value below $1\sigma$ will on the opposite indicate a poor precision.}
	\label{fig:ranking_nsigma}
\end{figure}

However, a fundamental information to bear in mind is also provided by the benchmark that also struggle sometime to reach a good score as for the 370-day signal on C800 or 187-day signal in B1300. This seems to indicate that the amount of information in the data was simply insufficient to reach a better level than 25\%. Recalling that such a benchmark was obtained using the ARVE solar red noise time-series without rotational modulation or instrumental stability consideration, this indicates that supergranulation alone when combined with suboptimal sampling can forbid such precision level. 

\begin{figure*}
	\centering
	\includegraphics[width=18.5cm]{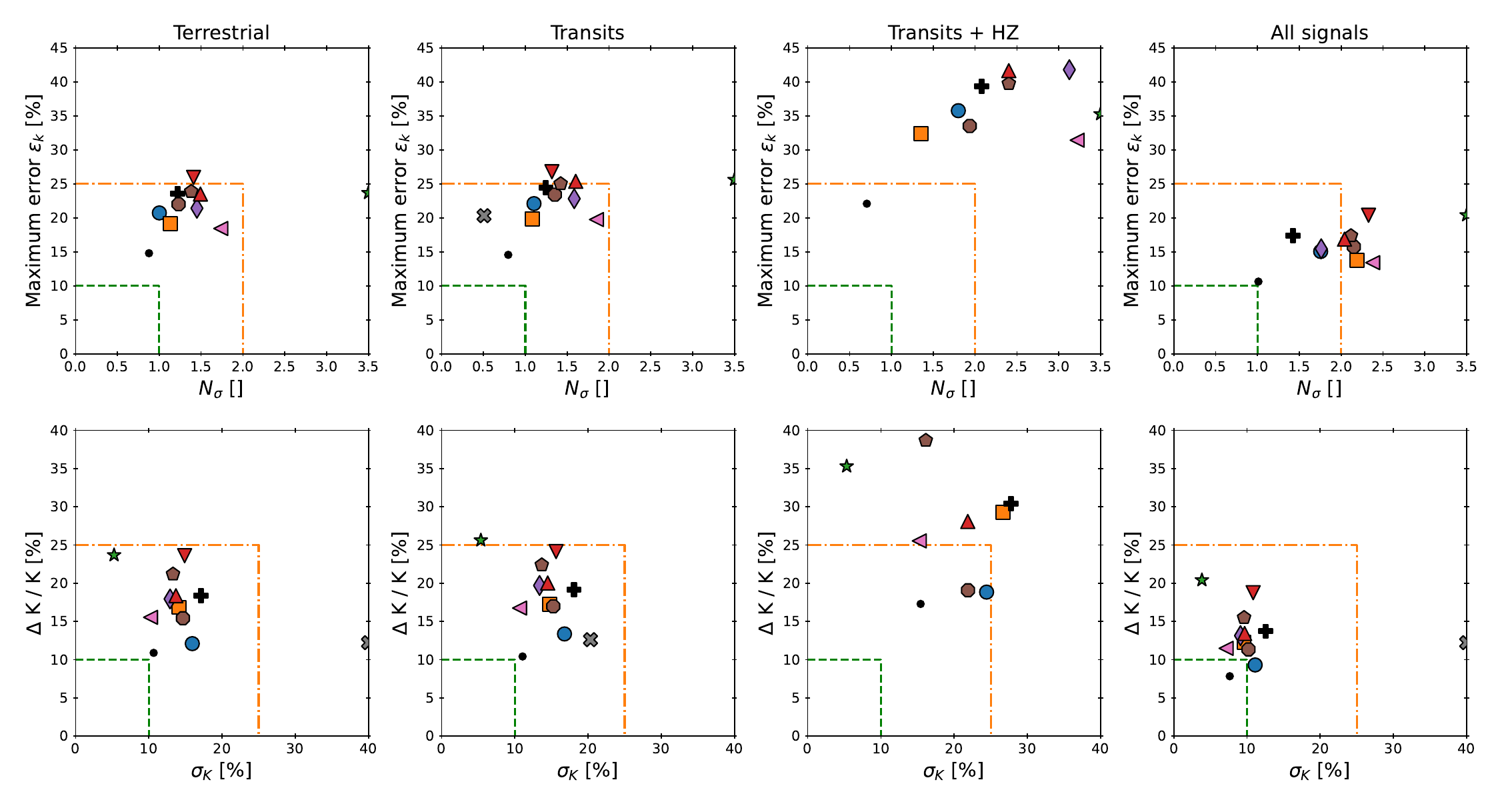}
	\caption{Representation of the four metrics \score\ $\varepsilon_K$, number of sigma $N_\sigma$, absolute \acc\ $|\Delta K / K|$ and \prec\ $\sigma_K$ obtained for each participants for the four different classes of signals. Top row represents the combination of ($N_\sigma$, $\varepsilon_k$) and bottom row represents the combination ($\sigma_K$, $\Delta K / K$) similarly to the previous Fig.\ref{fig:playground1}.}
	\label{fig:final_score}
\end{figure*}

We display in Fig.\ref{fig:participant_score} the \score\ $\varepsilon_K$ obtained by all the participants for each injected planet. The planets (columns) are ordered by the \score\ $\varepsilon_K$ of the benchmark. The participants (rows) are ordered according to their average $\varepsilon_K$. By carefully scrutinizing this figure, it becomes clear that the \score\ of the participants is in fact closely matching the benchmark even if slightly worse than it where the only exception is the dataset B1300.

\subsection{Ranking of the methods}
\label{sec:ranking}

There exists no good and unbiased ranking metric to order the participants. First because the ranking may depend on the type of signals, secondly because the ranking metric can be biased by systematic bias in the methodologies used by the participants (see below). We started by categorizing the injected signals into distinct classes. Four classes were defined to evaluate participant performance under different conditions: (1) terrestrial signals (2) transiting signals, (3) habitable-zone transiting signals, and (4) all the signals.

As an example of it, we computed for each class the average \score\ $\varepsilon_K$ of all participants\footnote{The grey cross corresponds to a partial submission of the challenge from the participant Sect.\ref{participant:tm}. Although it cannot be directly compared to the complete submissions, it is intentionally retained in the figure to highlight the potential risks of drawing conclusions from incomplete results.} and present the results in Fig.~\ref{fig:ranking_score}. Participants are ranked according to their \score\ evaluated over the "terrestrial signals" category (top panel), and this ordering is preserved across the remaining panels for consistency and ease of comparison. While the HZ transiting planets constitutes the primary focus of this challenge, it can not be used due to the low statistic number ($N=3$). The second best category of reference should have been the transiting planets, but due to the partial submission of one group, this metric is slightly biased for them and we therefore stuck to the terrestrial category for the reference. This choice is without consequence since all the transiting planets are terrestrial and as visible the ordering remains essentially the same using this group or the transiting one.

As expected, the best performance is achieved by the  benchmark participant described in Sect.~\ref{sec:white}. With a few exceptions, the ranking of the participants is largely insensitive to the signal category considered. The average \score\ $\varepsilon_K$ is slightly higher when all signals are included, since this subset contains the large gaseous planets (see Fig.~\ref{fig:KP_systems}), which are, by construction, the easiest signals to characterize.

No participant reaches a 10\% \score\ for the transiting signals, and three methods fall below 25\%: (1) YARARA V2, (2) Multi-GPs (Malavolta), and (3) Linear S-index (Damasso). It is worth noting that even the ARVE benchmark does not reach the 10\% level, indicating that such performance was simply not achievable for the injected signals given the number of observations and the adopted realistic noise level.

However, drawing conclusions solely from this global metric would be misleading. In Appendix~\ref{app:acc_prec}, we also present the rankings based separately on \prec\ $\sigma_K$ (Fig.~\ref{fig:ranking_precision}) and \acc\ $\Delta K /K$ (Fig.~\ref{fig:ranking_accuracy}). The precision metric is particularly useful for diagnosing methods that underestimate their uncertainties. Such cases can be identified when a method reports a precision better than the benchmark, which effectively represents a near best-case scenario.

\begin{figure*}
	\centering
	\includegraphics[width=14cm]{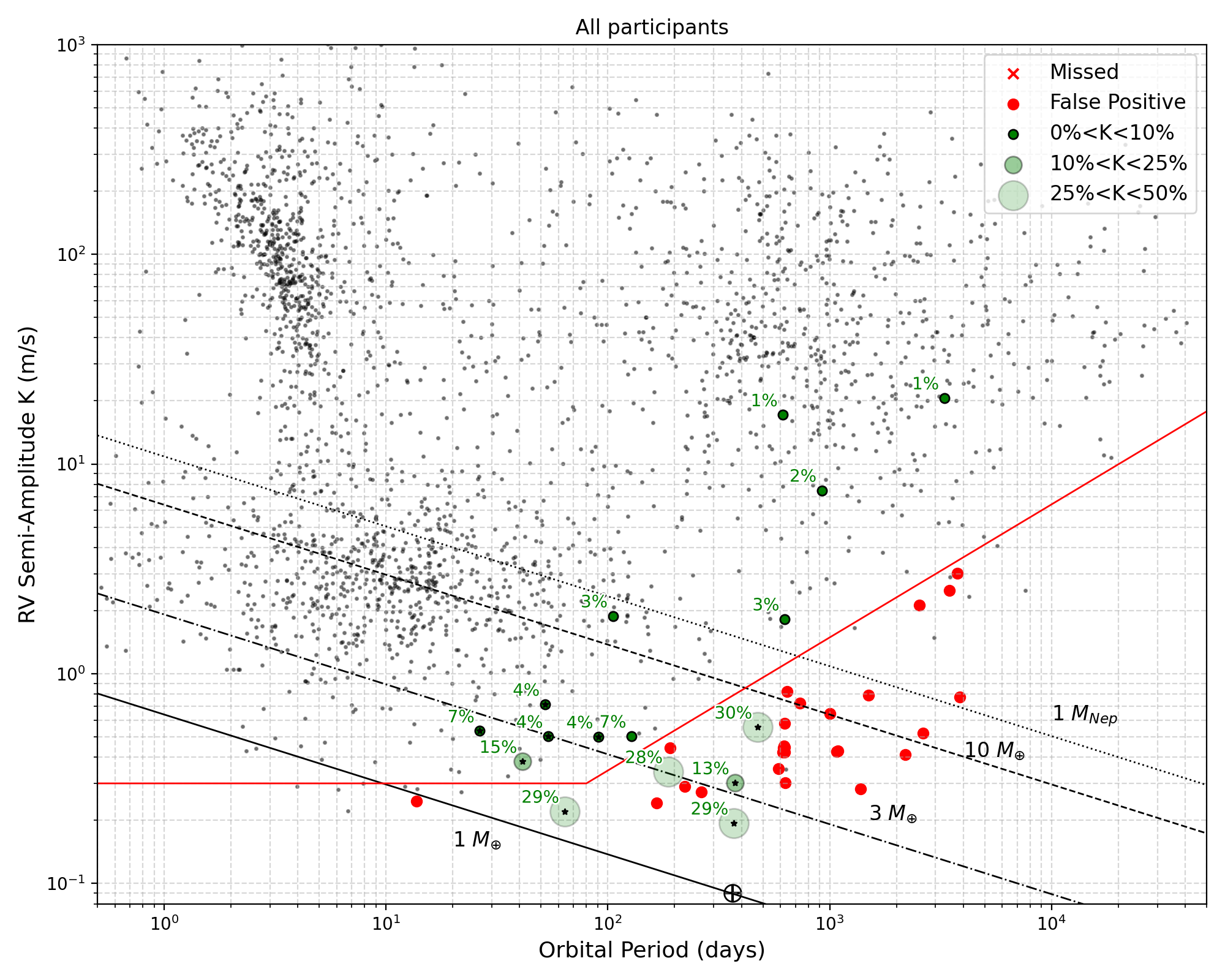}
	\caption{Visualization of the best score obtained by combining the results of all participants using the same ($K$,$P$) parameter space as the Fig.\ref{fig:KP_systems}. For each planet, only the best \score\ achieved among all participants was retained and is indicated by its value and the marker size (see the legend). As a counterpart, every false positive is also shown, even if detected by a single participant. All injected planetary signals were detected by at least one participant. The red lines delimit the ($K$, $P$) parameter-space region where the current RV precision enables reliable planetary detections from the region where published detections should be treated with caution.}
	\label{fig:best_score}
\end{figure*}

Two methods exhibit a better precision than the benchmark analysis (1) Shell CNN and (2) YARARA V2. This is suspicious, and likely indicates that the uncertainties are underestimated. A plausible explanation is that these methods do not simultaneously fit the Keplerian model and the red-noise component within a joint MCMC framework. Such rigid models might overlook the degeneracies between model components, and lead to underestimated uncertainties.

Systematic underestimation of uncertainties can bias the \score\ $\varepsilon_K$ ranking. Indeed, a method that overestimates its precision in a systematic way can achieve a better \score\ simply because the metric becomes dominated by its \acc\ while the other participants continue to be evaluated on both metrics. To enable a fair comparison, it is therefore essential to complement the \score\ analysis with an evaluation of the number-of-sigma compatibility $N_\sigma$ for each participant. This analysis is presented in Fig.~\ref{fig:ranking_nsigma}.

As visible by the comparison with the previous Fig.\ref{fig:ranking_score}, while YARARA V2 was among the best scoring methods, it is also among the methods that systematically overestimate its precision with one of the largest $N_\sigma$ deviation around $N_\sigma=1.5$ for terrestrial planets and even $N_\sigma\simeq3.0$ for HZ transiting ones.

In conclusion, we can not rely on a single one-dimensional ruler to decide of the success of a method and there always subsist a trade-off between accuracy and precision (or their alternative representation). We therefore display the average of each method in the 2D ($\Delta K /K$, $\sigma_K$) and ($\varepsilon_K$, $N_\sigma$) in the Fig.\ref{fig:final_score}. When analysing the methods that have both the smallest maximum error and are the most closely compatible to 1-sigma two methods can be highlighted that are the linear S-index detrending from Sect.\ref{participant:sindex2} and the multi-GP from Sect.\ref{participant:multigp}. Very interestingly, we see that the "best" participants themselves may depend if the metric used are the ($\varepsilon_K$, $N_\sigma$) from the top row or ($\Delta K /K$, $\sigma_K$) from the bottom row. 

In definitive, we do not clearly attest the existence of an outstanding method to the day and all methods seems to compete perfectly on such a data set. This is not extremely surprising given that some method may reveal their plain potential when working on more active stars, or on different SNR datasets.

\subsection{Assessing the current RV precision in the RV field}
\label{sec:rvfield}

Because this challenge involved a sufficiently large number of participants, employing a wide variety of methodologies — among the most innovative and representative currently available — we attempted to assess the present-day precision of the RV method.

The relatively small number of injected planetary signals is, of course, a limitation and should be kept in mind when interpreting the results, as should the caveats raised earlier regarding the extrapolation of results from the solar data set (see Sect.~\ref{sec:data set}).

It is becoming increasingly standard practice not to rely on a single pipeline or analysis when publishing planetary signals, but instead to run several independent pipelines in parallel and compare their outcomes. This healthy practice is now routine for JWST, where independent pipelines are executed in parallel during data processing \citep[e.g][]{Kirk(2024)}, and we advocate that a similar standard should progressively become the norm in RV studies.

However, combining results from different reductions or analyses remains a matter of debate. One cannot simply average the outputs of multiple pipelines, since they typically start from (or partially share) the same input data. This makes posterior propagation and model averaging non-trivial.

We therefore adopted a simpler approach: for each injected signal, we selected the best score achieved by any participant. Naturally, this choice introduces a bias and slightly overestimates the final \score \, $\varepsilon_k$. As a counterbalance, however, we treated all \emph{false positive} detections as valid detections, even if reported by only a single participant. By doing so, a statistic of 27 false detections is obtained by all the participants combined. The results of this analysis are shown in Fig.~\ref{fig:best_score}, and reveal an interesting trend.

Two distinct regions clearly emerge: one where detection scores are high and false positives are essentially absent, and another where scores are low and false detections\footnote{False detection seems "overrepresented" on the Fig.\ref{fig:best_score} that is a graphical bias since all participants produce the same green dots, but unique red dots.} become likely. Although the number of injected planets was insufficient to densely populate the $(K, P)$ parameter space, a rough empirical boundary $K_{\text{crit}}(P)$ can nevertheless be identified as follows for an HARPS-like instrument:

\begin{equation}
\label{eq:kcrit}
K_{\text{crit}}(P) =
\begin{cases}
30\,\mathrm{[cm\,s^{-1}]} & P \le 80 \, d\\
34 \left(\dfrac{P}{100\,\mathrm{d}}\right)^{0.634}\,\mathrm{[cm\,s^{-1}]} & P > 80 \, d
\end{cases}
\end{equation}

The number of false positives emerging from this challenge is a major concern, although they primarily occur — as expected — at low amplitudes and long orbital periods. This result may be concerning since a significant number of planets have been published in this region of the parameter space over the past years, Among the systems falling into the suspicious region are for instance: HD10700 f, HD20794 f, HD134060 c, HD95735 c, HD34445 g, GJ433 c and GJ15A c. 


Although we did not perform a case-by-case reassessment of those published detections, for instance to verify whether they were supported by independent methods, it is important to emphasize that our challenge was conducted under highly favorable conditions. The data sets typically contains an average of $N \sim 1000$ observations and were based on one of the most stable spectrographs currently available, operating under conditions and SNR that outperform most real stellar RV data sets obtained to date.

Given these considerations, our results suggest that caution is warranted when interpreting low-amplitude, long-period RV detections. In particular, we recommend that planets previously labeled as “confirmed” based solely on RV measurements, especially in this region of parameter space, should be reconsidered back as “candidates” unless supported by independent lines of evidence.

This further strengthens the case for the use of multiple independent pipelines and reduction strategies when assessing RV detections. Cross-validation between methods should become standard practice. However, our results also demonstrate that such sanity checks are not, by themselves, a guarantee against false positives.

In particular, among the false detections identified in the challenge, we find a $\sim 600$-day Super-Earth candidate ($\sim 5M_\oplus$) in data set B1300 that was independently detected by all participants. This illustrates that methodological diversity alone does not fully protect against systematic biases or activity-induced signals that can convincingly mimic planetary signatures.

\subsection{K-dwarf stars: the RV sweet spot?}
\label{sec:kdwarf}

In this section, we present a simple scaling analysis of the signals from the RV challenge to assess the impact of spectral type on the PLATO requirements.

Such an exercise should, however, be interpreted with caution. Changing the spectral type affects not only the stellar mass and the location of the HZ, but also the stellar activity budget and the spectral signatures. As a result, it is not straightforward to predict whether the various methods used by the participants would perform better or worse for different spectral types. Moving towards cooler spectral types nevertheless provides several advantages.

First, cooler stars naturally yield larger RV semi-amplitudes for a given planetary mass. In the HZ, the semi-amplitude scales as $K_{HZ} \propto M_{\star}^{-2/3}$ reflecting both the stellar-mass dependence and the inward shift of the HZ.

\begin{figure}
	\centering
	\includegraphics[width=8.4cm]{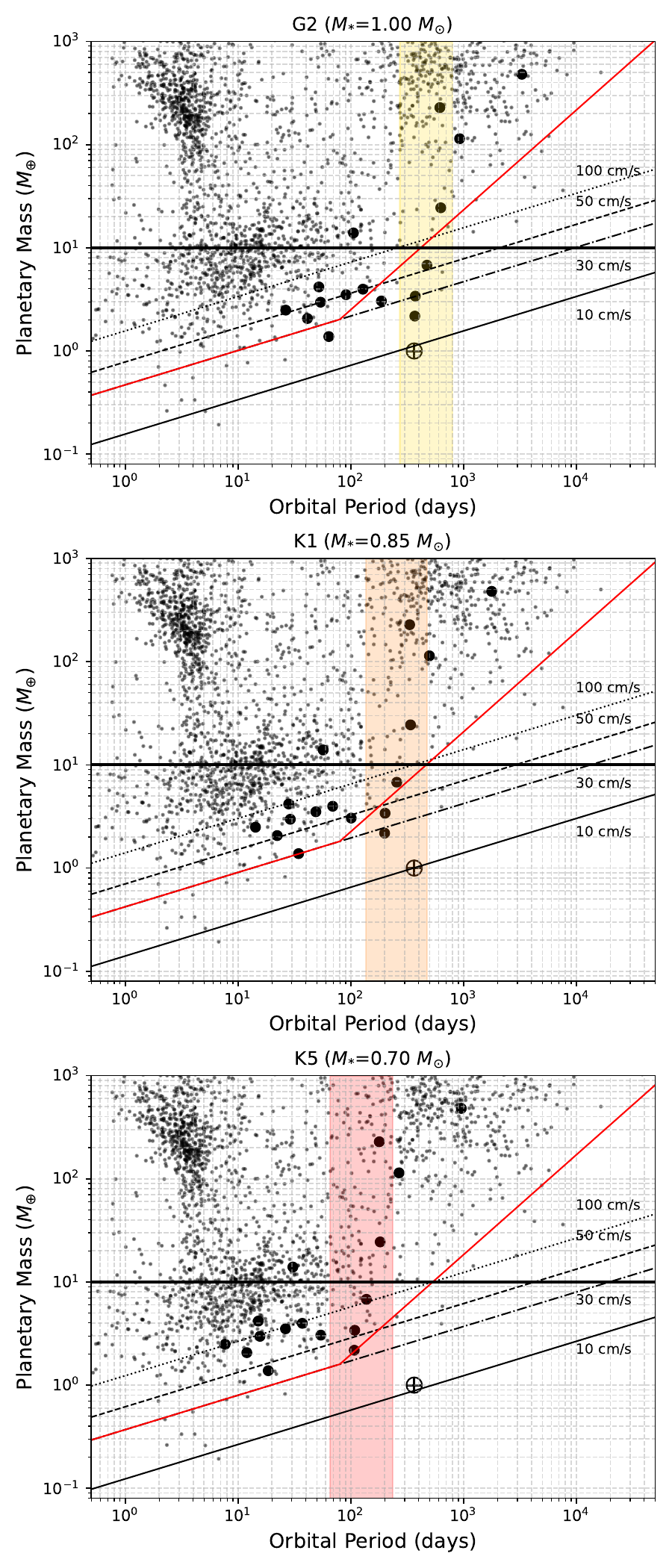}
	\caption{Scalability of the empirical detection limits from the PLATO RV challenge, derived for the Sun (top panel), and extrapolated to a K1 (middle panel) and K5 star (bottom panel) in the \textit{mass–period} diagram. The archival exoplanet database is displayed as comparison to the signals from this challenge (black markers). The HZ of each star is shown as a shaded region, computed using the conservative limits of \citet{Kopparapu(2013),Kopparapu(2014)}. The planetary masses from the RV challenge injected planets are kept fixed and periods rescaled to ensure Earth-Sun incident flux. For a G2 solar-type star, only planets with masses $m_p=10$ \Mearth \, (horizontal thick black line) satisfy the requirements at the inner edge of the HZ. The detection threshold improves for a K1 star, reaching $m_p=5 $\Mearth, and further improves significantly for a K5 star, where planets down to $m_p=2$ \Mearth \, reach confident detection level.}
	\label{fig:scalability}
\end{figure}

Second, the inward shift of the HZ leads to shorter orbital periods, which helps mitigate observational window effects linked to Earth’s orbit. This improves phase coverage and therefore enhances both detectability and characterization, as signals can be sampled over multiple orbital cycles.

Finally, the expected RV variability induced by convective flows is reduced. The amplitudes of p-modes, granulation (and supergranulation) are predicted to decrease by about 30\% \citep{Meunier(2019)}. In addition, cooler stars tend to rotate more slowly and exhibit lower spot contrast. As a consequence, stellar activity signals are expected to be more strongly dominated by the inhibition of convective blueshift (ICB) described in \citet{Meunier(2010)}, rather than by flux variations induced by active-region contrast \citep{Meunier(2013)}. The ICB amplitude being itself smaller since the convective flows are slower for cooler type-stars \citep{Liebing(2021), Almoulla(2024)}. 

We illustrate the expected gain in Fig.~\ref{fig:scalability}. We adopt the same KP detection limit as the red curve shown in Fig.~\ref{fig:final_score} and defined in Eq.~\ref{eq:kcrit}. This limit is then rescaled with stellar mass for two additional spectral types: K1 ($M_\star = 0.85,\Msun$) and K5 ($M_\star = 0.70,\Msun$). The HZ is correspondingly shifted inwards following \citet{Kopparapu(2013),Kopparapu(2014)}. The resulting HZ period ranges are $[272, 800]$ days for a G2 star, $[136, 472]$ days for a K1 star, and $[65, 237]$ days for a K5 star. 

We apply the same scaling to all planetary systems in the challenge by keeping the planetary masses fixed while rescaling their orbital periods such that HZ transiting planets receive the same incident flux as the Earth–Sun system. This analysis reveal that while for a G2 type star the requirement are achieved for $\sim$ 8 \Mearth\ planets for the innermost part of the HZ, the PLATO 10\% requirements may be fulfill down to $\sim$3 \Mearth\, for a K1 and $\sim$2 \Mearth\, for a K5 star. 

We note that similar scaling arguments have previously been used to motivate the search for HZ planets around M dwarfs, although several important limitations were not anticipated. In particular, M dwarf spectra in the visible are often heavily blended and obtained at relatively low signal-to-noise ratio, which has driven the development of infrared spectrographs. However, the technical challenges associated with infrared observations were initially underestimated \citep{Plavchan(2015),Fischer(2016)}, and their instrumental stability has only reach very recently the level achieved in the visible \citep{Artigau(2024),Bouchy(2025),Varas(2025)}.

Furthermore, M dwarfs typically host stronger magnetic fields, leading to enhanced stellar activity signals, in particular through Zeeman broadening \citep{Reiners(2013),Kochukhov(2021)}. Finally, the question of habitability itself has been rapidly called into question: the habitable zone lies very close to the host star, and young M dwarfs undergo intense flaring phases that may be sufficient to erode atmospheres or prevent the emergence of life.
\section{Conclusions}

\label{sec:conclusion}

In this paper, we presented an RV data challenge organized within the PLATO Stellar Variability Mitigation Working Group to compare different analysis methods and establish a robust analysis strategy in the context of the future PLATO follow-up, in particular regarding the precision and accuracy achievable on RV semi-amplitudes. 

The solar HARPS-N dataset was used to generate three synthetic datasets designed to closely mimic realistic follow-up observations. Three consecutive 5-minute exposures were combined to produce an effective integration time of 15 minutes, yielding datasets with $S/N_{550} \sim 540$. This signal-to-noise ratio corresponds to observations of a star with magnitude $m_v \approx 5.4$ using a 3.6m telescope such as HARPS, and $m_v \approx 7.2$ with ESPRESSO on the VLT.

Planetary signals were injected directly into the spectra, after which CCFs and RV time series were derived. Particular care was taken to ensure that the injected signals affected only the stellar component of the spectra, and not systematics tied to the Earth laboratory rest frame.

Three multi-planet systems were simulated, each containing between four and six planets, including one transiting habitable-zone planet as well as additional non-transiting companions. To limit the dimensionality of the problem, all planetary orbits were assumed to be circular, such that only the semi-amplitudes needed to be fitted for the known transiting planets. While this assumption simplifies the analysis, the impact of eccentricity on signal recovery could be explored in future studies.

Four metrics were used to evaluate participant performance based on the recovered K semi-amplitudes: (1) accuracy $\Delta K /K$, (2) precision $\sigma_k$, (3) maximum error $\varepsilon_k$, and (4) significance in units of sigma $N_\sigma$. No overall ranking was established, as performance was found to depend strongly on both the chosen metric and the specific signals. Moreover, the limited number of injected signals prevents any statistically robust ranking.

For terrestrial-mass signals, no participant achieved a precision better than 10\%, and even the benchmark time series reached only $\sim 15\%$. This indicates that supergranulation alone is sufficient to prevent reaching the PLATO requirement at this level. Most participants exhibit \score \, $\varepsilon_k$ in the range of 20–25\%.

By analysing both false positives and best-performing detections, a region of the $(K, P)$ parameter space was identified where planetary detections should be treated with caution for instruments with HARPS-like stability. Our results highlight the importance of comparing independent analysis methods when claiming low-amplitude planetary signals. In this regime, signals that are not consistently recovered by different methods should currently be considered as planet candidates rather than confirmed RV detections until corroborated by additional evidence.

Additional challenges arise from observational sampling, particularly for signals near a one-year period, which complicates the detection of HZ planets around solar-type (G2) stars. In contrast, for a similar magnitude, K dwarfs benefit from shorter HZ periods. Simple scaling relations suggest that the 10\% precision requirement could be met for HZ planets of $\sim 8$–$10,\Mearth$ around G2 stars, and down to $\sim 2$–$5,\Mearth$ around K dwarfs. These results further strengthen the case for K dwarfs as the most favourable targets for the first robust detections of habitable-zone planets with the RV technique. However, we noticed that this difficulty was enhanced in the RV challenge due to the differential extinction of the Sun producing season length of the three data sets around $S\sim 220$ days, while the visibility of the PLATO south field of view from Paranal is $S\sim 280$ days. 

This challenge provides three new benchmark data sets publicly available for the RV community to test and compare analysis methods. It should be noted, however, that future participants will be aware of the injected systems, whereas this challenge was conducted under near-blind conditions. 

Future extensions of this work include investigating the impact of lower SNR, eccentric signals and incorporating data from additional instruments such as ESPRESSO. The priority seems, to us, to improve the correction of granulation and supergranulation. The latter 
 appears to be one of the dominant limitations for future high-precision RV measurements. A dedicated data challenge focusing specifically on this component would be a natural next step. 

Some improvements on the planetary mass determination can be expected for the PLATO mission in particular if TTVs are used jointly to the Keplerian fit in RVs \citep{Leleu(2024)}, even if the improvement is yet unclear in presence of non-transiting planets. Best chance of improvements are expected if the supergranulation is better understood in the coming decade since this component alone is sufficient to make the 10\% difficult to achieve, in line with the results of \citet{Meunier(2020c)}. Such advancements are expected either thanks to PoET, the solar telescope of ESPRESSO \citep{Santos(2025)} or intensive future RV survey such as the second Earth project 2ES \citep{Sturmer(2024)} or the Terra Hunting Experiment \citep{Thompson(2016), Hall(2018)}  both entering in operation this year and having supergranulation as part of their open scientific question. 

In addition, the long-term stability of high-precision spectrographs should remain a priority, as instrumental interventions and ageing may introduce systematics that become critical on the multi-year timescales required to detect habitable-zone planets.

\section*{Acknowledgements}

MC and SA acknowledges
funding from the European Research Council (ERC) under the European Union’s Horizon 2020 research and innovation programme (grant agreement no. 865624). JBP, IR and MP were funded by Spanish grants PID2021-125627OB-C31,  PID2023-150491NB-I00, and PID2024-158486OB-C31, and by “ERDF A way of making Europe”,  by the programme Unidad de Excelencia María de Maeztu CEX2020-001058-M financed by MCIN/AEI/10.13039/501100011033 and by the MaX-CSIC Excellence Award MaX4-SOMMA-ICE, by the Generalitat de Catalunya/CERCA programme, and by the European Research Council (ERC) under the European Union’s Horizon Europe programme (ERC Advanced Grant SPOTLESS; no. 101140786). Views and opinions expressed are however those of the author(s) only and do not necessarily reflect those of the European Union or the European Research Council. Neither the European Union nor the granting authority can be held responsible for them. A.M. acknowledges funding from a UKRI Future Leader Fellowship, grant number MR/X033244/1 and a UK Science and Technology Facilities Council (STFC) small grant ST/Y002334/1. STH acknowledges support from UKRI-STFC grants ST/X001571/1 and UKRI974. ACC acknowledges support from STFC consolidated grant number ST/V000861/1, UKRI/ERC Synergy Grant EP/Z000181/1 (REVEAL), and UKSA grant number ST/X002217/1.
NCS is funded by the European Union (ERC, FIERCE, 101052347). Views and opinions expressed are however those of the author(s) only and do not necessarily reflect those of the European Union or the European Research Council. Neither the European Union nor the granting authority can be held responsible for them. 
This work was supported by Fundação para a Ciência e a Tecnologia (FCT) through national funds under the research grant UID/04434/2025 (DOI 10.54499/UID/04434/2025). The French researchers acknowledge support on PLATO by the Centre national d’études spatiales (CNES). TGW acknowledges support from the University of Warwick and UKSA. D.N. acknowledges support from PLATO ASI-INAF agreements no. 2022-28-HH.0. KA acknowledges support from the Swiss National Science Foundation (SNSF) under the Postdoc Mobility grant P500PT\_230225. NCS, KA, and EG acknowledge funding by the European Union (ERC, FIERCE, 101052347). Views and opinions expressed are however those of the author(s) only and do not necessarily reflect those of the European Union or the European Research Council. Neither the European Union nor the granting authority can be held responsible for them. This work was supported by Fundação para a Ciência e a Tecnologia (FCT) through national funds under the research grant UID/04434/2025 (DOI 10.54499/UID/04434/2025). C.A.W. would like to acknowledge support from the UK Science and Technology Facilities Council (STFC, grant number ST/X00094X/1).

This research has made use of NASA's Astrophysics Data System (ADS) bibliographic services. 
We acknowledge the community efforts devoted to the development of the following open-source packages that were used in this work: numpy (\href{http:\\numpy.org}{numpy.org}), matplotlib (\href{http:\\matplotlib.org}{matplotlib.org}), and astropy (\href{http:\\astropy.org}{astropy.org}). This publication makes use of The Data \& Analysis Center for Exoplanets (DACE), which is a facility based at the University of Geneva (CH) dedicated to extrasolar planets data visualisation, exchange and analysis. DACE is a platform of the Swiss National Centre of Competence in Research (NCCR) PlanetS, federating the Swiss expertise in Exoplanet research. The DACE platform is available at \url{https://dace.unige.ch}.

\section*{Data Availability}

The data underlying the RV challenge are publicly available in the Google Drive repository: \url{https://drive.google.com/drive/u/1/folders/1Jbz-i21D81Ilr90w3AVSDzygctGuFnqo}. Data are also publicly available on Zenodo (\url{10.5281/zenodo.20151946}).



\bibliographystyle{mnras}
\bibliography{mnras_template} 




\appendix

\section{Tables of participants output}

\begin{landscape}

\begin{table}
 \caption{Table of the $K_i$ semi-amplitudes posterior distributions of the participants for the planets in the challenge as well as false positive detections. The values represent the median of the distribution and the uncertainties estimated with 1.48 times the MAD. We indicated the false positive (FP) with their period in days in parenthesis. All the values for the K semi-amplitudes are in cm/s.}
 \label{tab:keplerians2}
\begin{tabular}{|c|c|lllllllllll|l|}
\hline
Planets  & Injected & \makecell{Exact\\Kep.\\(ARVE)} & \makecell{S-INDEX\\LINEAR\\(YVA)} & \makecell{Multi\\GPs\\(YVA)} & \makecell{SHELL\\CNN\\(YVA)} & \makecell{CCF\\SCALPELS\\(YVA)} & \makecell{CCF\\SCALPELS\\(YV0)} & \makecell{CCF\\LINSTAR\\(YVA)} & \makecell{S-INDEX\\LINEAR\\(YVA)} & \makecell{S-INDEX\\LINEAR\\(YV0)} & \makecell{YARARA\\YV2\\(YVA)} & \makecell{1D GP\\(YVA)} & \makecell{Participant\\average} \\
\hline\hline
A26d & $53$ & $55 \pm 3$ & $54 \pm 4$ & $57 \pm 4$ & $68 \pm 2$ & $54 \pm 3$ & $58 \pm 6$ & $58 \pm 4$ & $56 \pm 4$ & $54 \pm 5$ & $57 \pm 3$ &  & $56 \pm 6$ \\
A54d & $50$ & $53 \pm 3$ & $52 \pm 4$ & $51 \pm 4$ & $52 \pm 2$ & $54 \pm 4$ & $52 \pm 6$ & $54 \pm 4$ & $54 \pm 4$ & $54 \pm 5$ & $52 \pm 3$ &  & $52 \pm 5$ \\
A106d & $188$ & $195 \pm 3$ & $196 \pm 5$ & $194 \pm 4$ & $204 \pm 2$ & $193 \pm 5$ &  & $195 \pm 4$ & $196 \pm 4$ & $197 \pm 5$ & $197 \pm 3$ &  & $196 \pm 5$ \\
A373d & $30$ & $30 \pm 3$ & $33 \pm 4$ & $38 \pm 4$ & $38 \pm 2$ & $42 \pm 3$ & $39 \pm 5$ & $40 \pm 4$ & $33 \pm 4$ & $34 \pm 5$ & $35 \pm 3$ &  & $36 \pm 7$ \\
A627d & $182$ & $183 \pm 4$ & $193 \pm 4$ & $191 \pm 4$ & $223 \pm 2$ & $196 \pm 6$ & $224 \pm 9$ & $195 \pm 5$ & $189 \pm 5$ & $187 \pm 5$ & $197 \pm 4$ &  & $193 \pm 8$ \\
A922d & $749$ & $757 \pm 4$ & $775 \pm 4$ & $770 \pm 4$ & $823 \pm 2$ & $768 \pm 4$ & $767 \pm 8$ & $772 \pm 7$ & $769 \pm 5$ & $767 \pm 5$ & $765 \pm 4$ &  & $770 \pm 7$ \\\hline
FP1 &  &  & 41 (2185d) &  & 29 (222d) &  & 250 (3465d) & 52 (2626d) &  &  &  &  &  \\ 
\hline\hline
B41d & $38$ & $44 \pm 3$ & $48 \pm 5$ & $45 \pm 4$ & $45 \pm 2$ & $48 \pm 4$ & $49 \pm 4$ & $46 \pm 5$ & $48 \pm 5$ & $49 \pm 5$ & $44 \pm 3$ & $49 \pm 11$ & $46 \pm 4$ \\
B52d & $71$ & $72 \pm 3$ & $75 \pm 4$ & $77 \pm 4$ & $74 \pm 2$ & $74 \pm 4$ & $74 \pm 4$ & $73 \pm 5$ & $75 \pm 5$ & $75 \pm 5$ & $75 \pm 3$ & $74 \pm 8$ & $74 \pm 4$ \\
B91d & $50$ & $57 \pm 4$ & $51 \pm 5$ & $50 \pm 4$ & $52 \pm 2$ & $54 \pm 5$ & $50 \pm 5$ & $51 \pm 5$ & $50 \pm 5$ & $51 \pm 5$ & $50 \pm 3$ & $53 \pm 10$ & $51 \pm 5$ \\
B128d & $50$ & $51 \pm 4$ & $51 \pm 4$ & $57 \pm 5$ & $54 \pm 2$ & $52 \pm 3$ & $60 \pm 5$ & $51 \pm 4$ & $51 \pm 5$ & $52 \pm 5$ & $53 \pm 3$ & $56 \pm 66$ & $51 \pm 6$ \\
B187d & $34$ & $42 \pm 4$ &  &  &  &  &  &  &  & $44 \pm 5$ &  &  & $44 \pm 5$ \\
B474d & $56$ & $57 \pm 4$ & $80 \pm 5$ &  & $80 \pm 2$ & $80 \pm 7$ &  & $80 \pm 5$ & $80 \pm 5$ & $72 \pm 5$ & $83 \pm 4$ &  & $80 \pm 5$ \\ \hline
FP1 &  &  & 44 (620d) & 25 (14d) & 42 (627d) & 42 (612d) & 82 (642d) & 45 (621d) & 45 (624d) &  58 (626d) & 30 (630d) & 43 (1073) &  \\
FP2 &  &  &  & 35 (586d) &  & 28 (1381d) & 212 (2528d) & 24 (166d) &  &  &  &  &  \\
FP3 &  &  &  & 79 (1498d) &  &  & 301 (3768d) &  &  &  &  &  &  \\
FP4 &  &  &  & 77 (3860d) &  &  &  &  &  &  &  &  &  \\
\hline\hline
C64d & $22$ & $23 \pm 4$ & $28 \pm 7$ & $32 \pm 5$ & $39 \pm 2$ & $33 \pm 6$ & $33 \pm 6$ & $25 \pm 6$ & $33 \pm 6$ & $31 \pm 6$ & $32 \pm 4$ &  & $32 \pm 7$ \\
C370d & $19$ & $29 \pm 6$ & $20 \pm 10$ & $26 \pm 8$ &  & $20 \pm 8$ & $32 \pm 6$ & $29 \pm 5$ & $20 \pm 8$ & $34 \pm 5$ & $22 \pm 6$ &  & $25 \pm 10$ \\
C616d & $1714$ & $1718 \pm 4$ & $1754 \pm 11$ & $1755 \pm 5$ & $1919 \pm 2$ & $1742 \pm 8$ & $1750 \pm 7$ & $1744 \pm 14$ & $1755 \pm 6$ & $1759 \pm 5$ & $1733 \pm 5$ &  & $1752 \pm 13$ \\
C3286d & $2063$ & $2070 \pm 6$ & $2114 \pm 26$ & $2213 \pm 34$ &  & $2139 \pm 20$ & $2142 \pm 16$ & $2115 \pm 27$ & $2134 \pm 13$ & $2120 \pm 12$ & $2081 \pm 6$ &  & $2107 \pm 86$ \\ \hline
FP1 &  &  &  & 27 (264d) & 44 (190d) &  &  & 43 (1090d) &  &  & 64 (1000d) &  &  \\
FP2 &  &  &  &  & 72 (733d) &  &  &  &  &  &  &  &  \\
\hline
\end{tabular}
\end{table}

\end{landscape}

\section{Evaluation of the spectra RV shift injection on the CCFs moments}
\label{app:injection}

A major update of this RV challenge compared to previous efforts is the injection of planetary signals directly into the spectra. To implement this, we first evaluated the impact of the RV-shift injection algorithm in order to validate its use for the challenge.

\begin{figure*}
	\centering
	\includegraphics[width=18cm]{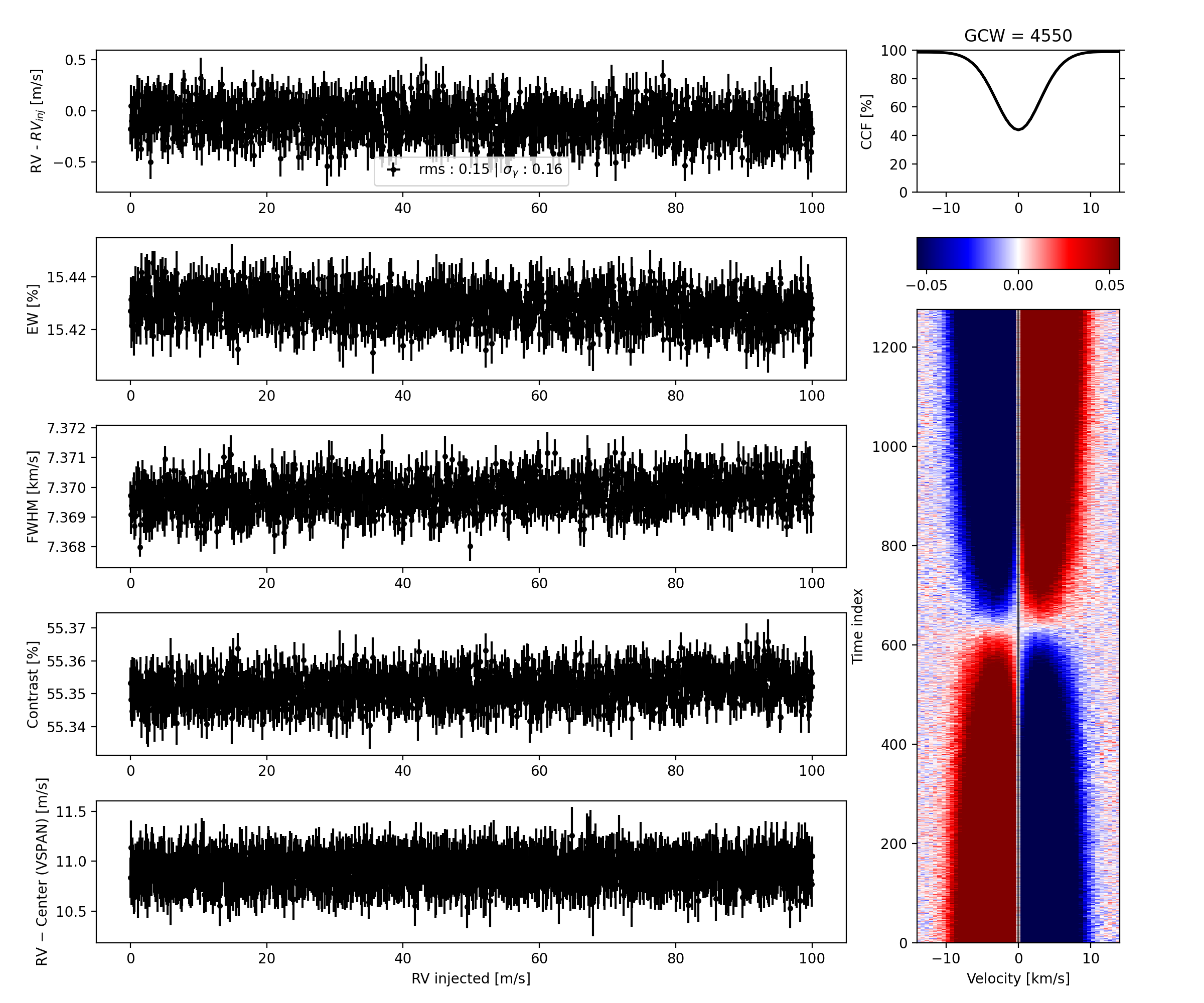}
	\caption{Evaluation of the RV shift injection method used in this RV challenge. The same spectrum was duplicated and shifted on a grid of RV shift between 0 to 100 m/s. Photon noise injection were done independently on each spectra to mimic the $\sigma_{RV}\simeq$15 cm/s RV precision of the challenge. The CCFs are displayed in the top right corner. The residual of the CCFs with the median is shown in the right heat map while the different CCF moments as fitted from a Gaussian fit are displayed on the different horizontal panels. No significant deviation is observed between the injected RVs and the recovered ones for all the tested velocity shifts. Uncertainties are the propagated photonic flux uncertainties and match the intrinsic dispersion. A small leak of the RV shift interpolation is visible in the FWHM (third panel) as visible by the small linear trend. However restricting the RV shift to 20 m/s }
	\label{fig:test1}
\end{figure*}

In principle, the cleanest way to inject a planetary signal is to apply a Doppler shift to the wavelength solution prior to computing the CCFs using the standard extraction algorithm. However, this approach was not compatible with our framework, as the RV challenge relies on YARARA to process and clean the spectral time series.

YARARA operates on a spectral time-series matrix in which all spectra share a common wavelength grid. This uniform grid is a core assumption of the data format required by the code. Consequently, any injected RV shift must be followed by a reinterpolation step back onto the native wavelength grid.

While reinterpolation is sometimes regarded with caution in high-resolution spectroscopy, there is no strong evidence that it significantly degrades the data quality when applied carefully. In the worst case, interpolation may introduce a slight increase in the noise level, but this effect is typically negligible and does not materially affect the scientific interpretation.

In our implementation, the shifted spectra were reinterpolated onto the original grid using cubic splines. Naturally, the validity of such an interpolation depends on the amplitude of the Doppler shift. What constitutes a “small” shift must therefore be evaluated in the specific context of the challenge.

To quantify the impact of the method, we selected a representative spectrum from the data set and created an exact copy. Photon noise was then injected to reproduce the final RV precision of the solar data set used in the challenge, namely $\sigma_{RV}\simeq 15$ cm/s. We applied Doppler shifts ranging from 0 to 100 m/s, followed by reinterpolation, exactly as performed in the construction of the RV challenge.

The results are shown in Fig.~\ref{fig:test1}. We did not observe any significant deviation between the injected and recovered RV values over the full tested range up to 100 m/s. However, a weak linear trend is visible in the full width at half maximum (FWHM) of the CCF, indicating a small leakage of the Doppler shift into the line-profile shape.

We quantified this leakage by fitting a linear trend to the CCF moments. For the RV itself, we obtained:

$$RV - RV_{inj} = -0.172 \times RV_{inj}\, \left[\frac{cm/s}{m/s}\right]  $$

This implies that for an injected shift of 100 m/s, the recovered RV is underestimated by 17 cm/s. This is a relative difference of 0.172\% on the injected amplitude, far smaller than the 10\% objective of this challenge. Such a bias is extremely small which demonstrate that the method selected here will not introduce any significant bias for the participants and confirm the validity of the method. However, given the precision requirements of the challenge, we therefore limited the Keplerian semi-amplitudes to 20 m/s, ensuring that the maximum bias remains below 3.4 cm/s.

Furthermore, restricting the injected amplitudes to this range also prevents significant contamination of the CCF moments, since over such RV ranges the leakage remains below the intrinsic jitter precision of the data set for all considered CCF diagnostics.

\section{Ranking of the participants with  precision and accuracy}
\label{app:acc_prec}

While an attempt was made to develop a single one-dimensional ruler to rank the participants, for instance by the use of the \score\ $\varepsilon_k$ metric that combines both the \acc\ and \prec\, we realise that all the score metric are unperfect to some degree and still require a 2D parameters space to analyse the results. 

\begin{figure}
	\centering
	\includegraphics[width=8cm]{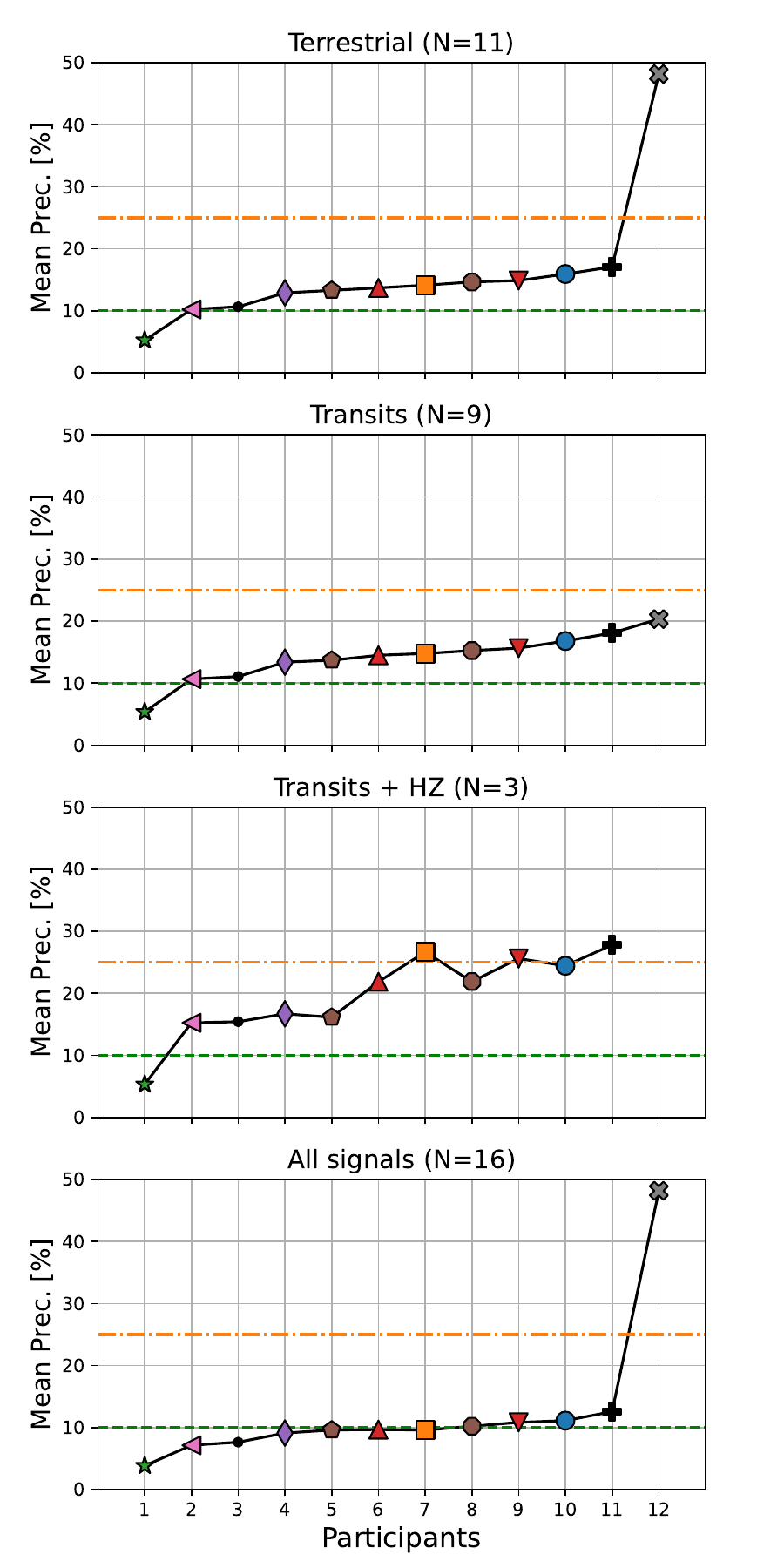}
	\caption{Same as Fig.\ref{fig:ranking_score} for the \prec\ $\sigma_K$ instead of \score\ $\varepsilon_K$. In general it can be see that the \prec\ reached by all the participants is lower than their \acc\ (Fig.\ref{fig:ranking_accuracy}), meaning that all the participants tend to slightly overestimate their \prec\ in comparison to their \acc. Furthermore, participants with a \prec\, better than the white benchmark (black dots ranked in $3^{\text{rd}}$) are certainly underestimating their uncertainties when fitting the Keplerian model. }
	\label{fig:ranking_precision}
\end{figure}

\begin{figure}
	\centering
	\includegraphics[width=8cm]{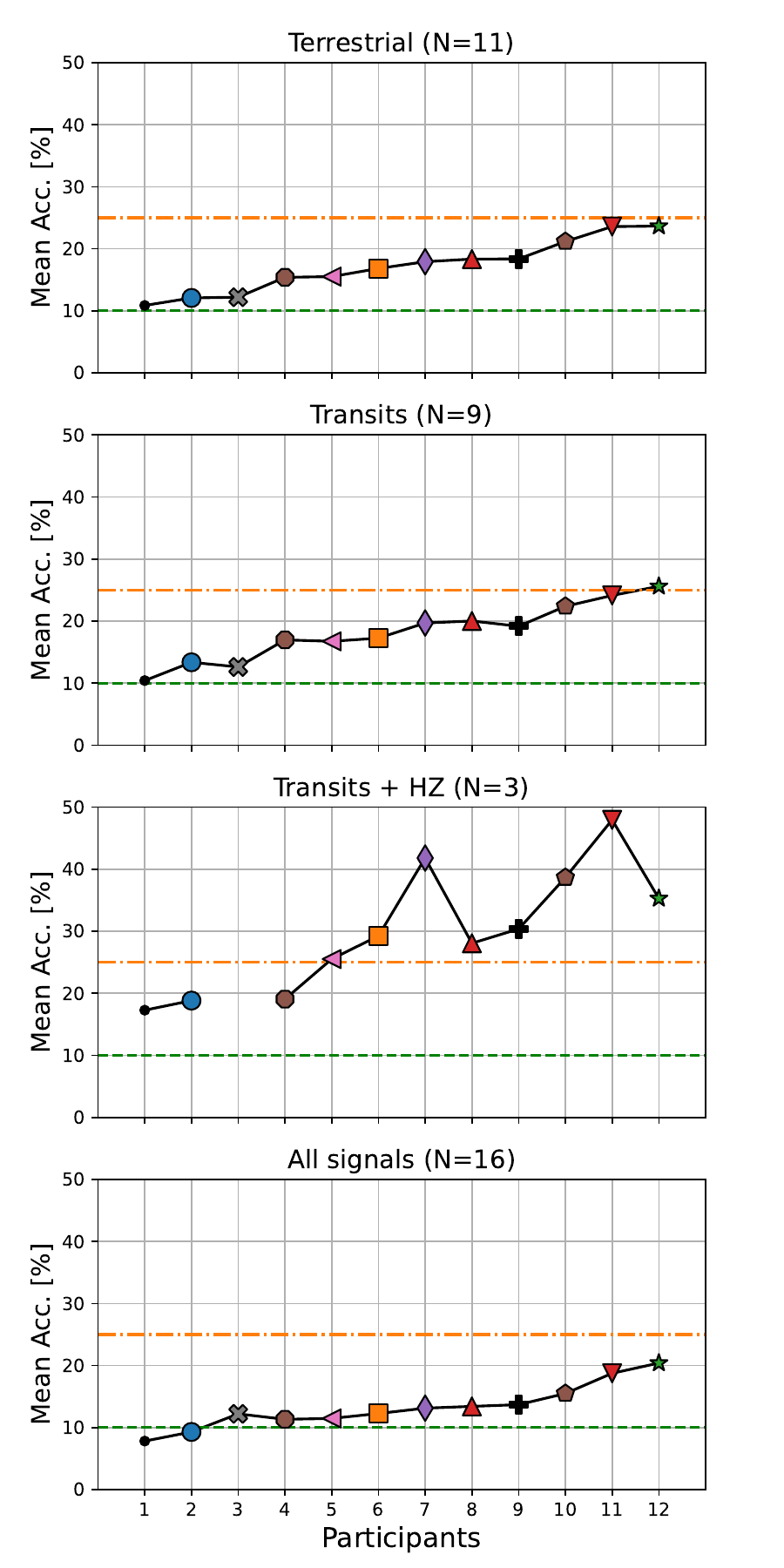}
	\caption{Same as Fig.\ref{fig:ranking_score} for the \acc\ $\Delta K/K$ instead of \score\ $\varepsilon_K$. The best accuracy reached on the transiting planet by the benchmark was expected to be around 15 \%.}
	\label{fig:ranking_accuracy}
\end{figure}

We display the ranking of all the participants based on each individual metric, respectively the \acc\ in Fig.\ref{fig:ranking_accuracy} and the \prec\ in Fig.\ref{fig:ranking_precision}.

\bsp	
\label{lastpage}
\end{document}